\documentclass[12pt]{article}

\newcommand{\asm}{\alpha_s(m_{\tau})} 

\usepackage{citesort}

\usepackage{graphicx}

\newcommand{\M}[3]{{\mathrm{T}}^{}_{#2#3}}

\newcommand{\ice}[1]{\relax}

\ice{

\setlength{\textwidth}{16cm}
\setlength{\textheight}{22.5cm}
\setlength{\oddsidemargin}{1.cm}
\setlength{\headsep}{1.0cm}
\setlength{\topmargin}{0.0cm}
\setlength{\topskip}{0.1cm}
\setlength{\footskip}{1.5cm}
}
\newcommand{\ep}{\epsilon}

\newcommand{\prt}{\partial }

\newcommand{\GeV}{\mbox{GeV}}

\newcommand{\beq}{\begin{equation}}
\newcommand{\eeq}{\end{equation}}
\newcommand{\bea}{\begin{eqnarray}}
\newcommand{\eea}{\end{eqnarray}}

\newcommand{\ba}{\begin{array}} 
\newcommand{\ea}{\end{array}} 
 
\newcommand{\als}{\alpha_s}
\newcommand{\as}{a_s}

\newcommand{\asp}[1]{a^{\prime #1}_s}

\newcommand{\alsnf}{\alpha_s^{(n_f)}}
\newcommand{\alsnl}{\alpha_s^{(n_l)}}

\newcommand{\g}{\gamma}

\newcommand{\msbar}{\overline{\mbox{MS}}}

\newcommand{\dsp}{\displaystyle}

\newcommand{\EQN}{\label}
\newcommand{\ovl}{\overline}
\newcommand{\re}[1]{(\ref{#1})}

\newcommand{\BreakI}{ \right. \nonumber \\ &{}& \left. }

\newcommand{\Lc}{{\cal L}}

\def\bbuildrel#1_#2^#3%
{\mathrel{\mathop{\kern 0pt#1}\limits_{#2}^{#3}}}

\newcommand{\n}{\hspace*{-2.5mm}}

\begin{document}

\begin{titlepage}
\noindent

{

\centerline{\normalsize\hfill  SFB/CPP-05-80}
\centerline{\normalsize\hfill  TTP05-27     }
\baselineskip 11pt
{}
}

\vspace{0.5cm}
\begin{center}
  \begin{Large}
  \begin{bf}
QCD Decoupling \ice{and On-Shell EM  Charge Renormalization} at Four Loops
 \end{bf}
  \end{Large}

s%
 \vspace{0.8cm}

  \begin{large}
    K.G. Chetyrkin${}^{a,}$\footnote{On leave from Institute for
      Nuclear Research of the Russian Academy of Sciences, Moscow, 117312,
      Russia.}, J.H.~K\"uhn${}^a$, 
    C.~Sturm${}^b$ {\normalsize  } 
  \end{large}
  \vskip .7cm

         {\small {\em
             ${}^a$Institut f\"ur Theoretische Teilchenphysik,
             Universit\"at Karlsruhe,
             D-76128 Karlsruhe, Germany
\\
${}^b$ Dipartimento di Fisica Teorica, Universit{\`a} di Torino, 
Italy \& INFN, Sezione di Torino, Italy
}}

        \vspace{0.8cm}
{\bf Abstract}
\end{center}
\begin{quotation}
\noindent
We present the matching condition for the strong coupling contant
$\alpha_s^{(n_f)}$ at a heavy  quark  threshold to four loops in the
modified minimal subtraction scheme.  Our results lead to further
decrease of the theoretical uncertainty of the evolution of the strong 
 coupling constant through heavy quark thresholds.  Using a low 
energy theorem we  furthermore derive the effective coupling of the Higgs boson
to gluons (induced  by a virtual  heavy quark) in  four-
and (partially) through five-loop approximation. 

\end{quotation}
\end{titlepage}

\section{Introduction \label{sec:intro}}

The masses of the known quark species differ vastly in their
magnitude. As a result of this, in many QCD applications the mass of a
heavy quark $h$ is much larger than the characteristic momentum scale
$\sqrt{s}$ which is intrinsic to the physical process. In such a
situation there appear two interrelated problems when using an  MS-like
renormalization scheme.

First, given the two large but in general quite different mass scales,
$\sqrt{s}$ and $m$, two different types of potentially
dangerously large logarithms may arise. The standard trick of a proper choice of
the renormalization scale $\mu$ is no longer effective; one can not set
one parameter $\mu$ equal to two very different mass scales simultaneously. 

Second, according to the Appelquist-Carazzone theorem
\cite{Appelquist:1974tg} 
the effects due to heavy particles 
 should eventually  
``decouple'' from the low-energy physics\footnote{It should be stressed
that the statement is literally valid only if power-suppressed
corrections of order $(s/m^2)^n$ with $n>0$ are neglected.}. 
However, a peculiarity of mass-independent renormalization schemes is that the
decoupling theorem does not hold in its naive form for theories
renormalized in this  framework. The effective QCD action to appear will
not be canonically normalized. Even worse, potentially large mass logarithms in
general appear, when one calculates a physical observable. \\

The well-known way to  bring  the large mass logarithms under control is to
construct an effective field theory by ``integrating out'' the heavy
quark field $h$ 
\cite{Weinberg:1980wa,Ovrut:1980dg,Wetzel:1981qg,%
Bernreuther:1981sg,Bernreuther:1983zp}.
By construction, the resulting effective QCD
Lagrangian will not include the heavy quark field. In order to be
specific, let us  consider QCD with $n_l=n_f-1$ light quarks and one
heavy quark $h$ with mass $m$. 
The quark-gluon  coupling constants in both theories, the full $n_f$-flavor QCD
and the effective $n_l$-flavor one,  
$\alsnf$ and $\alsnl$ are related by the so-called 
matching condition of the form \cite{Weinberg:1980wa,Ovrut:1980dg}
\begin{equation}
\alsnl(\mu)=
\zeta_g^2 (\mu,\alsnf(\mu),m)\,\,
\alsnf(\mu)
{}.
\label{als_matching_cond}
\end{equation}
Here ${m} = m(\mu)$ is the (running) mass of the heavy quark; 
the decoupling  function 
$\zeta_g$ is equal to one at the leading (tree) level but receives  nontrivial
corrections in higher orders. 
The matching point $\mu$ in  eq.~\re{als_matching_cond} should be chosen
in   such a  way to minimize the effects of  logarithms of the heavy quark mass,
e.g. $\mu = {\cal O}(m)$.

An important phenomenological application of 
eq.~\re{als_matching_cond} appears in  the determination of $\alpha_s(M_Z)$ at
the $Z$-boson scale through evolution  with the renormalization group
equation, starting from the measured value  $\alpha_s(m_{\tau})$ at the
$\tau$-lepton scale. A careful analysis of the effects of four-loop
running and three-loop matching in extracting $\alpha_s^{(5)}(M_Z)$ with 5
active quark flavors from $\alpha_s^{(3)}(m_{\tau})$ with 3 active
quark flavors has been recently performed in
ref.~\cite{Davier:2005xq}. We will demonstrate that the inclusion of the
newly computed four-loop matching condition leads to further reduction
of the theoretical error from the evolution.

The second aspect, mentioned above, is of importance for Higgs-boson
production in hadronic collisions.
The dominant subprocess for the production of the Standard-Model (SM)
Higgs boson at the CERN Large Hadron Collider (LHC) will be the one
via gluon fusion. Therefore, an important ingredient for the
Higgs-boson search will be the effective coupling of the Higgs boson
to gluons, usually called $C_1$.  In ref.~\cite{Chetyrkin:1998un} a
low-energy theorem, valid to all orders, was established, which 
relates the effective Higgs-boson-gluon coupling, induced by the
virtual presence of a heavy quark, to the logarithmic derivative of
$\zeta_g$ with respect to the heavy quark mass.  
In that paper the method was used 
to squeeze from the three-loop decoupling function $\zeta_g$ 
the analytical result for $C_1$ in three- and even in four-loop approximation 
(the latter in  a indirect way through a sophisticated use of the RG evolution equations
and the four-loop QCD $\beta$-function).  With our new full four-loop
result for $\zeta_g$ we are able to confirm the result of ref.~\cite{Chetyrkin:1998un}
in a completely independent way, without using the four-loop contribution to the
QCD $\beta$-function. In addition a five-loop prediction for $C_1$ can
be obtained (modulo yet unknown contribution from the five-loop
$n_f$-dependent term in the $\beta$-function).

The outline of the paper is as follows.  In
Section~\ref{sec:formalism}  we recall the main formulae from
\cite{Chetyrkin:1998un} which reduce the evaluation of the decoupling
function $\zeta_g$ to the calculation of vacuum integrals.  Then we
discuss shortly the technique used to compute these integrals.
Section 3 describes our four-loop results for the decoupling function.
In Section 4 we make use of the low energy theorem to derive the
effective coupling of the Higgs boson to gluons (due to the virtual
presence of a heavy quark) through four and (partially) through five
loops. Section 5 deals with phenomenological applications of our
result which lead to a reduction of  the theoretical
error due to evolution of the coupling constant $\alpha_s$ through
heavy quark thresholds. Summary and conclusions are presented in the
final Section 6.

\section{Formalism\label{sec:formalism}}
\subsection{Decoupling for the Gauge Coupling Constant \label{subsec:dec}}
Let us consider QCD with $n_{l} = n_{{f}} -1$  massless  quarks
$\psi =
\{
\psi_{{l}}| l= 1 \mbox{-} n_{l}
\}
$
and one  heavy quark $h$ with mass $m$.
The corresponding  bare QCD  Lagrangian  reads 
\beq
\ba{c}
\dsp
\Lc(g_0,m_0,\xi_0;\psi_0, G^{0,a}_\mu, c^{0,a},h_0)=-\frac{1}{4} (G^{0,a}_{\mu\nu})^2 +
i\ovl\psi_0 \!\not{\!\! D} {\psi_0} + \ovl h_0 (i\!\not{\!\! D_0} - m_0) h_0
\\
\dsp
+ \ \mbox{\rm terms with ghost fields and the gauge-fixing term},
\ea
\EQN{lgr0}
\eeq
where  the index `0' marks bare quantities, 
$G^{0,a}_\mu $ and 
$c^{0,a}$ are  the gluon and ghost field respectively.
The relations between the bare and renormalized quantities read
\begin{eqnarray}
g_s^0&\n=\n&\mu^{\varepsilon}Z_gg_s,\qquad
m^0=Z_m m,\qquad
\xi^0-1=Z_3(\xi-1),
\nonumber\\
\psi_q^0&\n=\n&\sqrt{Z_2}\psi_q,\qquad
G_\mu^{0,a}=\sqrt{Z_3}G_\mu^a,\qquad
c^{0,a}=\sqrt{\tilde{Z}_3}c^a,
\label{eqren}
\end{eqnarray}
where $g_s=\sqrt{4\pi\alpha_s}$ is the (renormalized) QCD gauge coupling,
$\mu$ is the renormalization scale,
$d=4-2\varepsilon$ is the dimensionality of space-time.
The gauge parameter $\xi$ is defined through the gluon propagator in lowest
order,
\begin{equation}
\frac{i}{q^2+i\epsilon}\left(-g^{\mu\nu}+\xi\frac{q^\mu q^\nu}{q^2}\right)
\label{gauge:cov}
{}.
\end{equation}
All renormalization constants appearing in (\ref{eqren}) are 
known  by now through order ${\cal O}(\alpha_s^4)$
from 
\cite{vanRitbergen:1997va,Vermaseren:1997fq,Chetyrkin:1997dh,Chetyrkin:2004mf,Czakon:2004bu}.

Integrating out the heavy quark transforms the full QCD
Lagrangian  (\ref{lgr0}) into the one corresponding to the effective massless
QCD with $n_f' = n_f-1 =n_l$ quark flavors (plus additional higher
dimension interaction terms  suppressed by powers of the heavy mass and 
neglected in  what follows).  
Denoting the effective fields and parameters
by an extra prime,  we write the effective Lagrangian as follows:
\begin{equation}
\Lc^{\prime}={\cal L}\left(g_s^{0\prime},\xi^{0\prime};
\psi_q^{0\prime},G_\mu^{0\prime,a},c^{0\prime,a}\right)
{}.
\end{equation}
Here the primed quantities are related to non-primed ones through 
\begin{eqnarray}
g_s^{0\prime}&\n=\n&\zeta_g^0 g_s^0,\qquad
\xi^{0\prime}-1=\zeta_3^0(\xi^0-1),
\qquad 
\psi_q^{0\prime}=\sqrt{\zeta_2^0}\psi_q^0,\qquad
\nonumber\\
G_\mu^{0\prime,a}&\n=\n&\sqrt{\zeta_3^0}G_\mu^{0,a},\qquad
c^{0\prime,a}=\sqrt{\tilde\zeta_3^0}c^{0,a},
\label{eqdec}
\end{eqnarray}
where the primes mark the quantities of the effective $n_l$-flavor theory.

As was first demonstrated in \cite{Chetyrkin:1998un} 
the bare decoupling constants can all be expressed in terms of 
massive Feynman integrals without any external momenta (so-called massive tadpoles). 
The corresponding relation for $\zeta_g^0$ reads:
\begin{equation}
\zeta_g^0=\frac{\tilde\zeta_1^0}{\tilde\zeta_3^0\sqrt{\zeta_3^0}}
\label{zetag0}
{},
\end{equation}
where
\begin{eqnarray}
\tilde\zeta_1^0&\n=\n& 1+\Gamma_{G\bar cc}^{0h}(0,0),
\\
\zeta_3^0&\n=\n&1+\Pi_G^{0h}(0),
\\
\tilde\zeta_3^0&\n=\n&1+\Pi_c^{0h}(0)
{}.
\label{eqdeccon2}
\end{eqnarray}
Here 
$\Pi_G(p^2)$ and
$\Pi_c(p^2)$ are the gluon and ghost vacuum polarization functions, respectively.
$\Pi_G(p^2)$ and $\Pi_c(p^2)$ are related to the gluon and ghost
propagators through
\begin{eqnarray}
&\n{}\n&
\hspace*{-4cm}\delta^{ab}
\left\{\frac{g^{\mu\nu}}{p^2\left[1+\Pi_G^0(p^2)\right]}
+\mbox{terms proportional to $p^\mu p^\nu$}\right\}
\\
&=&i\int dx\,e^{ip\cdot x}
\left\langle TG^{0,a\mu}(x)G^{0,b\nu}(0)\right\rangle,
\nonumber\\
-\frac{\delta^{ab}}{p^2\left[1+\Pi_c^0(p^2)\right]}
&\n=\n&i\int dx\,e^{ip\cdot x}
\left\langle Tc^{0,a}(x)\bar{c}^{0,b}(0)\right\rangle,
\end{eqnarray}
respectively.
The vertex function  $\Gamma_{G\bar cc}^0(p,k)$ is defined through the
one-particle-irreducible (1PI) part of the amputated $G\bar cc$ Green function
as
\begin{eqnarray}
\lefteqn{p^\mu g_s^0\left\{-if^{abc}\left[1+\Gamma_{G\bar cc}^0(p,k)\right]
+\mbox{other color structures}\right\}}
\nonumber\\
&\n=\n&i^2\int dxdy\,e^{i(p\cdot x+k\cdot y)}
\left\langle Tc^{0,a}(x)\bar c^{0,b}(0)G^{0,c\mu}(y)\right\rangle^{\rm 1PI},
\end{eqnarray}
where $p$ and $k$ are the outgoing four-momenta of $c$ and $G$, respectively,
and $f^{abc}$ are the structure constants of the QCD gauge group.

Finally, in order to renormalize the bare decoupling constant $\zeta_g^0$ we
combine eq.~\re{zetag0} with eqs.~\re{eqren} and arrive at
($\alpha_s = g_s^2/(4\pi), \alpha^{\prime}_s = (g^{\prime}_s)^2/(4\pi)$)
\beq
\alpha_s^\prime(\mu) =
\left(\frac{Z_g}{Z_g^\prime}\zeta_g^0\right)^2 \, \alpha_s(\mu)
=\zeta_g^2\alpha_s(\mu)
{}.
\label{eqdecg}
\eeq
Direct application of this equation is rather clumsy as the
constant ${Z_g^\prime}$ on its right side depends on the
{\em renormalized} effective coupling constant which we are looking for.
Of course, within perturbation theory, one can always solve eq.~\re{eqdecg} by
iteration. A simpler way, which we have used, reduces to inverting first the 
series 
\beq
\alpha^0_s = Z_{\alpha}(\alpha_s,\ep) \,\alpha_s, \ \ Z_{\alpha}(\alpha_s,\ep)
=  Z_g^2= 1+ \sum_{i \ge 1} Z_{\alpha,i}(\ep) \, \alpha_s^i
\ \ 
\eeq
to express the {\em renormalized} coupling constant $\alpha_s$ in terms of the
{\em bare one}:
\beq
\alpha_s = Z^0_{\alpha}(\alpha^0_s,\ep) \,\alpha^0_s, \ \ Z^0_{\alpha}(\alpha_s,\ep)
 = 1+ \sum_{i \ge 1} Z^0_{\alpha,i}(\ep) \, \left(\alpha^0_s \right)^i
{},
\eeq
where, up to four loops, 
\bea
Z^0_{\alpha,1} &=& -Z_{\alpha,1}, \ \  Z^0_{\alpha,2} = -Z_{\alpha,2} +2\, Z_{\alpha,1}^2
{}, 
\ \
Z^0_{\alpha,3} = -Z_{\alpha,3} +5 \, Z_{\alpha,1}\, Z_{\alpha,2} -5 \, Z_{\alpha,1}^3
\nonumber
\\
Z^0_{\alpha,4} &=& -Z_{\alpha,4} +6 \, Z_{\alpha,1} \, Z_{\alpha,3} +3\, Z_{\alpha,2}^2 
- 21\,Z^2_{\alpha,1}\, Z_{\alpha,2} +14  \, Z_{\alpha,1}^4
\label{Zg0}
{}\,.
\eea
\ice{
Out[62]//InputForm= 
1 - asp*Z1 + asp^2*(2*Z1^2 - Z2) + asp^3*(-5*Z1^3 + 5*Z1*Z2 - Z3) + 
 asp^4*(14*Z1^4 - 21*Z1^2*Z2 + 3*Z2^2 + 6*Z1*Z3 - z4)
}
With the use of eq.~\re{Zg0} one now could conveniently transform all the
primed (that is effective quantities ) in eq.~\re{eqdecg} into the
non-primed ones. After this the renormalization can be done directly.

\subsection{Vacuum Integrals \label{subsec:vac}}

At the end of the day we have to evaluate  a host of four-loop  massive tadpoles   
entering the  definitions of the bare decoupling constants.
The evaluation of these massive tadpoles in three-loop approximation
has been pioneered in ref. \cite{Broadhurst:1991fi} and automated in
ref. \cite{Steinhauser:2000ry}.\\
Similar to the three-loop case, the analytical evaluation of four-loop tadpole
integrals is based on the traditional  Integration-By-Parts (IBP) method.
In contrast to the three-loop case the manual
construction of algorithms to reduce arbitrary diagrams to a small set
of master integrals is replaced by Laporta's algorithm
\cite{Laporta:1996mq,Laporta:2001dd}.
In this context the IBP identities are generated
with numerical values for the powers of the propagators and the
irreducible scalar products. In  the next step, the resulting system of linear
equations is  solved  by expressing systematically complicated
integrals in terms of simpler ones. The resulting  solutions are then
substituted into all the other equations.\\
This reduction has been implemented in an automated {\tt{FORM3}}
\cite{Vermaseren:2000nd,Vermaseren:2002rp} based program in which
partially ideas described in
ref. \cite{Laporta:2001dd,Mastrolia:2000va,Schroder:2002re} have been
implemented. The rational functions in the space-time dimension $d$, which
arise in this procedure,  are simplified with the program {\tt{FERMAT}}
\cite{Lewis}. The automated exploitation of {\em all} symmetries of the
diagrams by reshuffling the powers of the propagators of a given
topology in a unique way strongly reduces  the number of
equations which need to be solved.

In general, the tadpole diagrams contributing to the decoupling constants 
contain both massive and massless lines. In contrast, the
computation of the four-loop $\beta$-function can be reduced to the
evaluation  of four-loop tadpoles  composed of {completely
massive} propagators.  These special cases have been
considered  in \cite{vanRitbergen:1997va,Schroder:2002re,Czakon:2004bu}.

All four-loop tadpole diagrams encountered during our calculations
were expressed through the set of 13 master integrals shown in figures
1 and 2. The first four integrals of Fig. 1 have analytical solution
in terms of Gamma functions for generic valuse of $\ep$. The first
nine terms of the $\ep$-expansion of the fifth integral ($T_{52}$) can
be obtained from results of Refs.~
\cite{Broadhurst:1991fi,Broadhurst:1996az} (for more details see
Ref.~\cite{Schroder:2005va}). As for the less simple master integrals
pictured in Fig. 2 one finds that for the case of $\zeta_g$ all
necessary ingredients are known analytically
(\cite{Laporta:2002pg,Chetyrkin:2004fq,Schroder:2005db,Kniehl:2005yc,%
Schroder:2005va,Chetyrkin:2005}) except for $T_{54,3}$,
$T_{62,2}$ and $T_{91,0}$, where we have denoted
\beq
{T_{n}}\, \cdot\,\left(
\ep \cdot \int\frac{ \mathrm{d}^{4-2\ep} p}{\pi^{2-\ep}}\frac{1}{(p^2 +1)^2}
\right)^{(-4)}
\bbuildrel{=\!=\!=}_{\ep \to 0}^{} \,\,  \sum_{i \, \ge\,  -4} \ep^i \, T_{n,i}
\,
{}.
\eeq
These  three integrals are known  numerically  from 
\cite{Schroder:2005va,Chetyrkin:2005} and read
\bea
T_{54,3} &=& -8445.8046390310298, 
\\
T_{62,2} &=& -4553.4004372195263, 
\\
T_{91,0} &=&      1.808879546208
{}.
\eea
In fact, in refs.~\cite{YS_MS,Chetyrkin:2005} an analytical result for
$T_{91,0}$ (in terms of $T_{62,2}$ and $T_{54,3}$) has been also
derived. Thus, the results of the next section will contain only {\em
two} numerical constants, viz.  $T_{62,2}$ and $T_{54,3}$.

\begin{figure}[!ht]
\begin{center}
\begin{minipage}[b]{2cm}
  \begin{center}
    \includegraphics[height=1.5cm,bb=126 332 460 665]{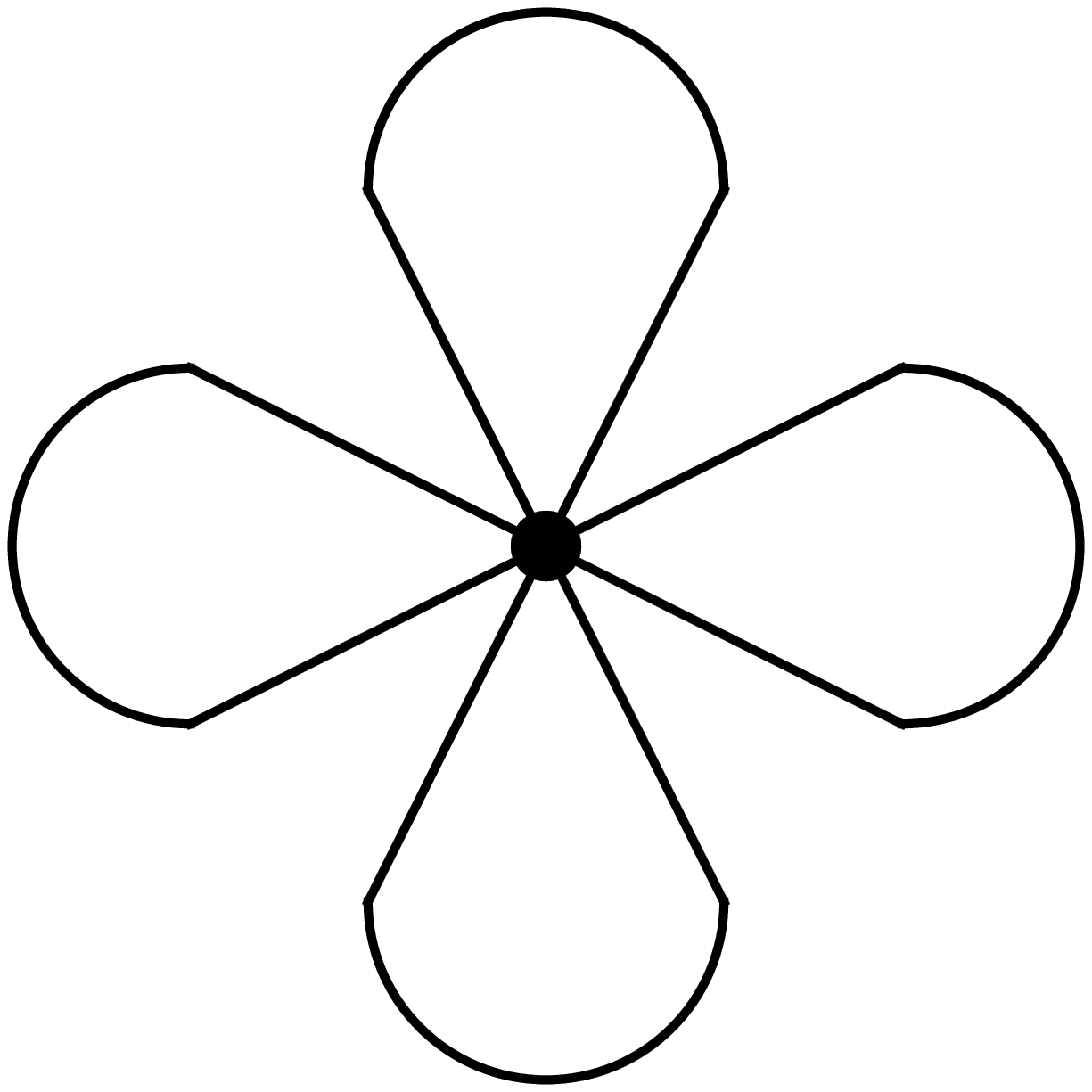}
    $\M{}{4}{1}$
  \end{center}
\end{minipage}
%
%
\begin{minipage}[b]{2cm}
  \begin{center}
    \includegraphics[height=1.5cm,bb=170 320 415 666]{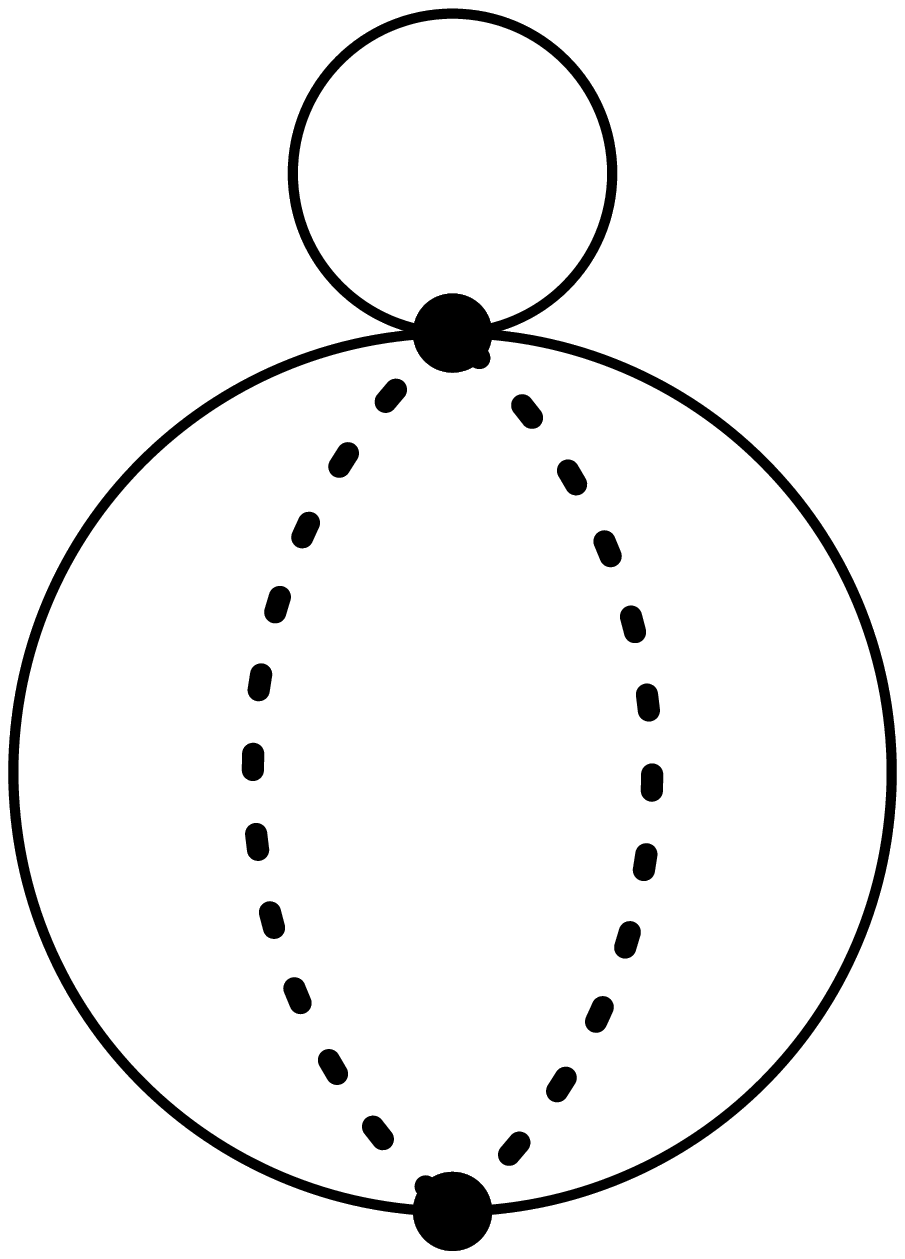}
\\
    $\M{}{5}{1}$
  \end{center}
\end{minipage}
%
%
\begin{minipage}[b]{2cm}
  \begin{center}
    \includegraphics[height=1.5cm,bb=170 320 415 666]{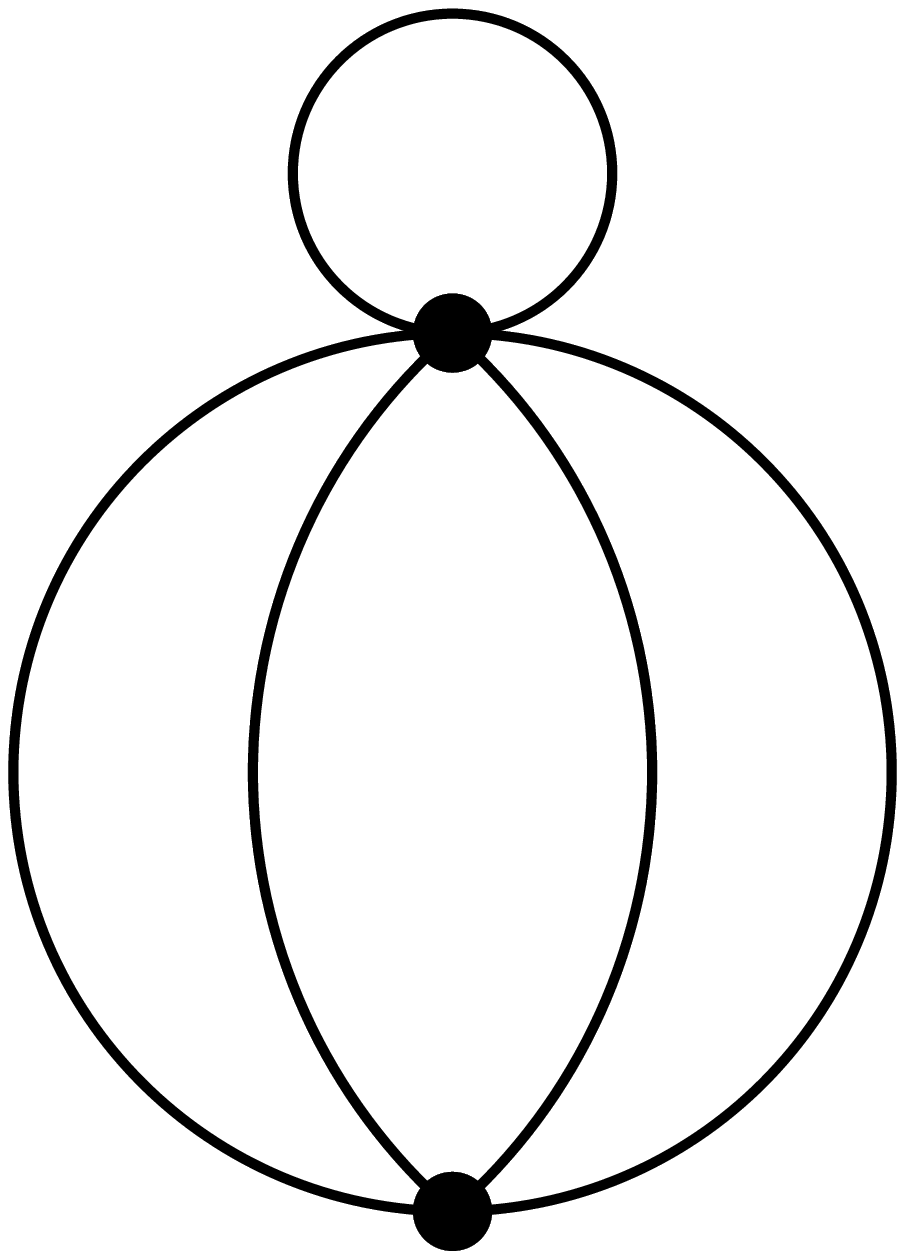}
\\
    $\M{}{5}{2}$
  \end{center}
\end{minipage}
%
%
\begin{minipage}[b]{2cm}
  \begin{center}
    \includegraphics[height=1.5cm,bb=126 320 460 678]{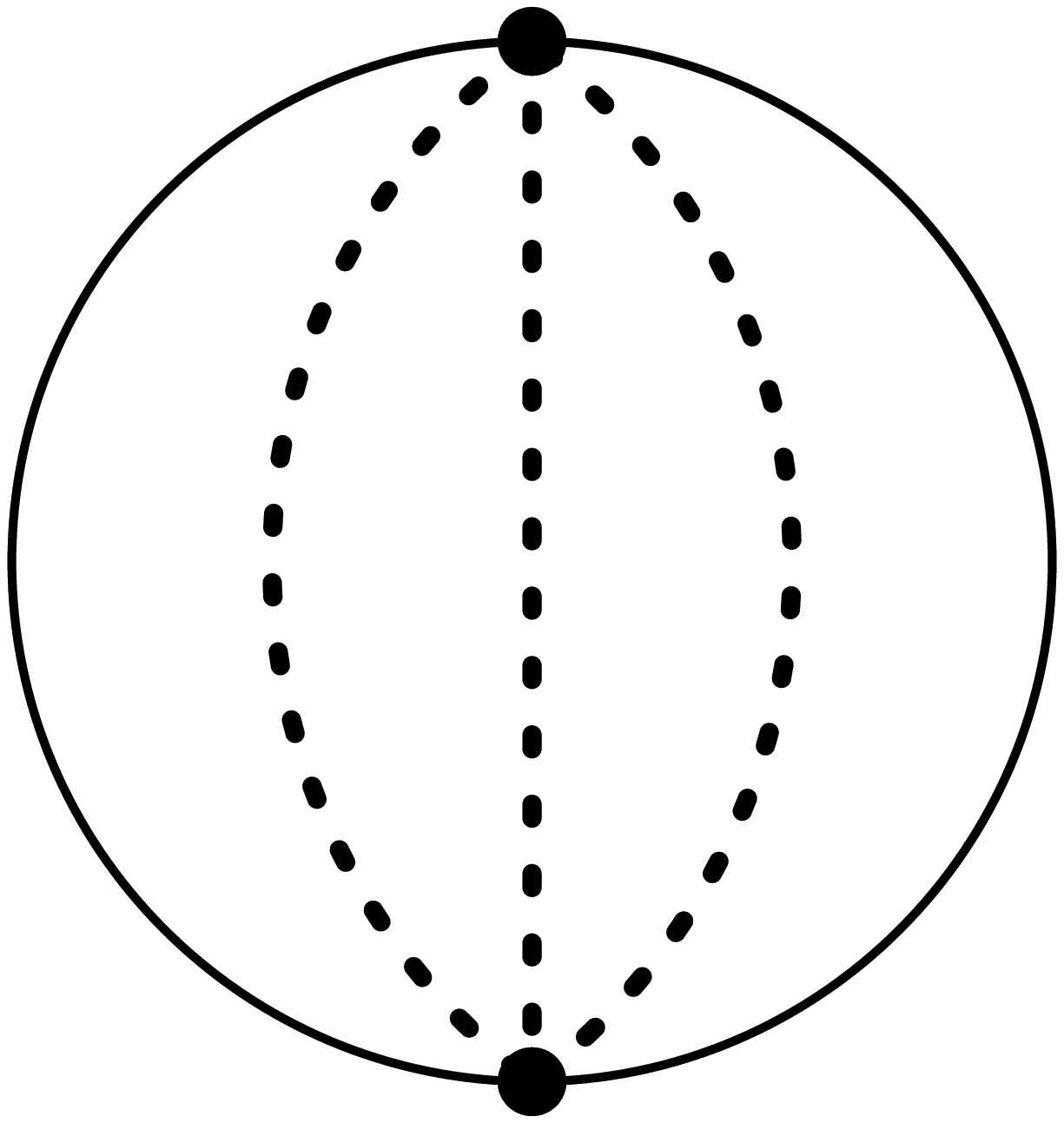}
    $\M{}{5}{3}$
  \end{center}
\end{minipage}
%
%
\begin{minipage}[b]{2cm}
  \begin{center}
    \includegraphics[height=1.5cm,bb=126 320 460 678]{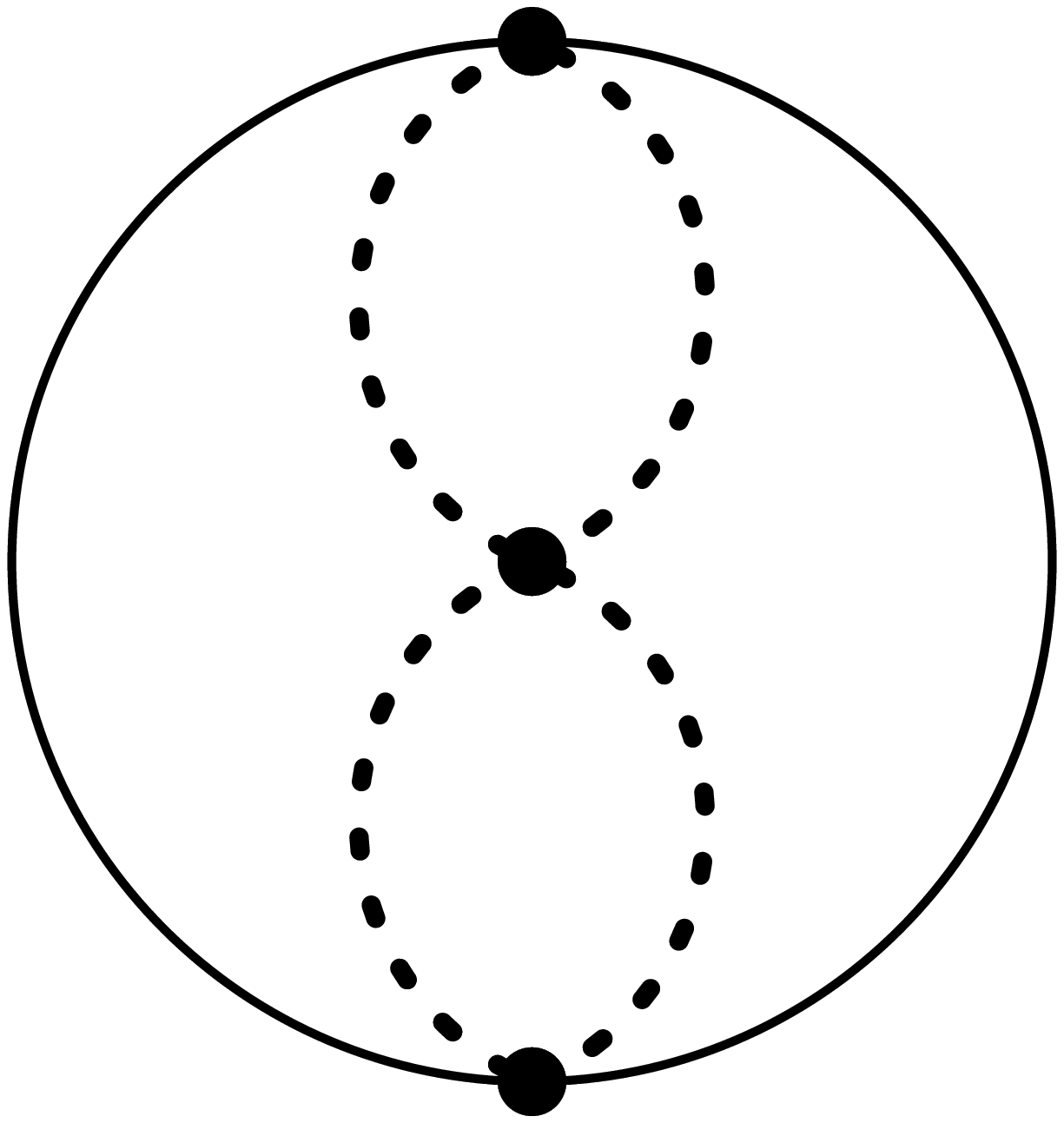}
    $\M{}{6}{3}$
  \end{center}
\end{minipage}
 \end{center}
\caption{
Analytically known  master integrals 
\label{masters:simple}} 
\end{figure}

%
\begin{figure}[!ht]
 \begin{center}
\begin{minipage}[b]{3.0cm}
  \begin{center}
    \includegraphics[height=1.5cm,bb=126 320 460 678]{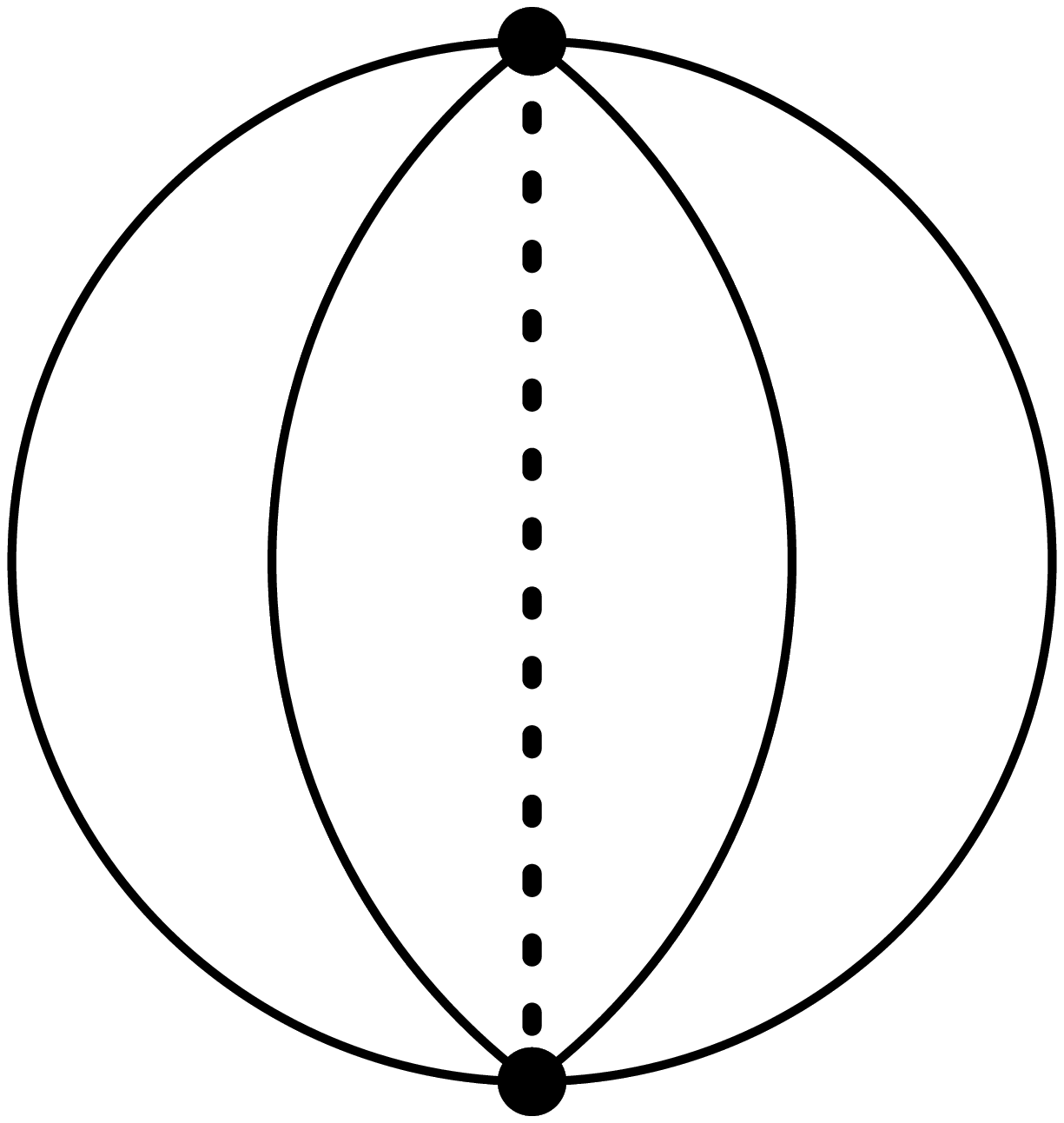}
    $\M{}{5}{4}, (4,3,2)$
  \end{center}
\end{minipage}
%
\begin{minipage}[b]{3.0cm}
  \begin{center}
    \includegraphics[height=1.5cm,bb=126 320 460 678]{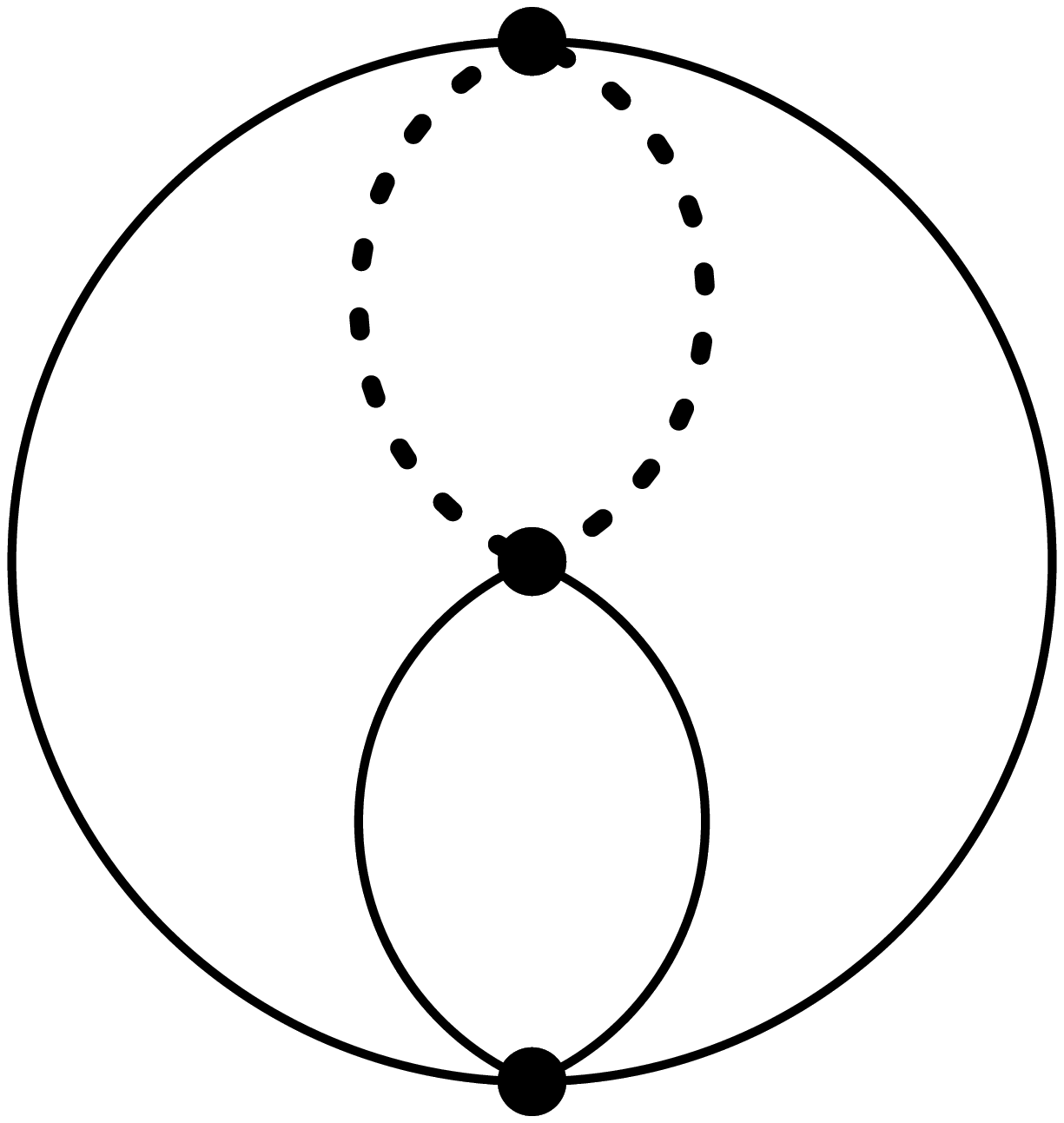}
    $\M{}{6}{2}, (3,2,1) $
  \end{center}
\end{minipage}
%
%
\begin{minipage}[b]{3.0cm}
  \begin{center}
    \includegraphics[height=1.5cm,bb=126 320 460 678]{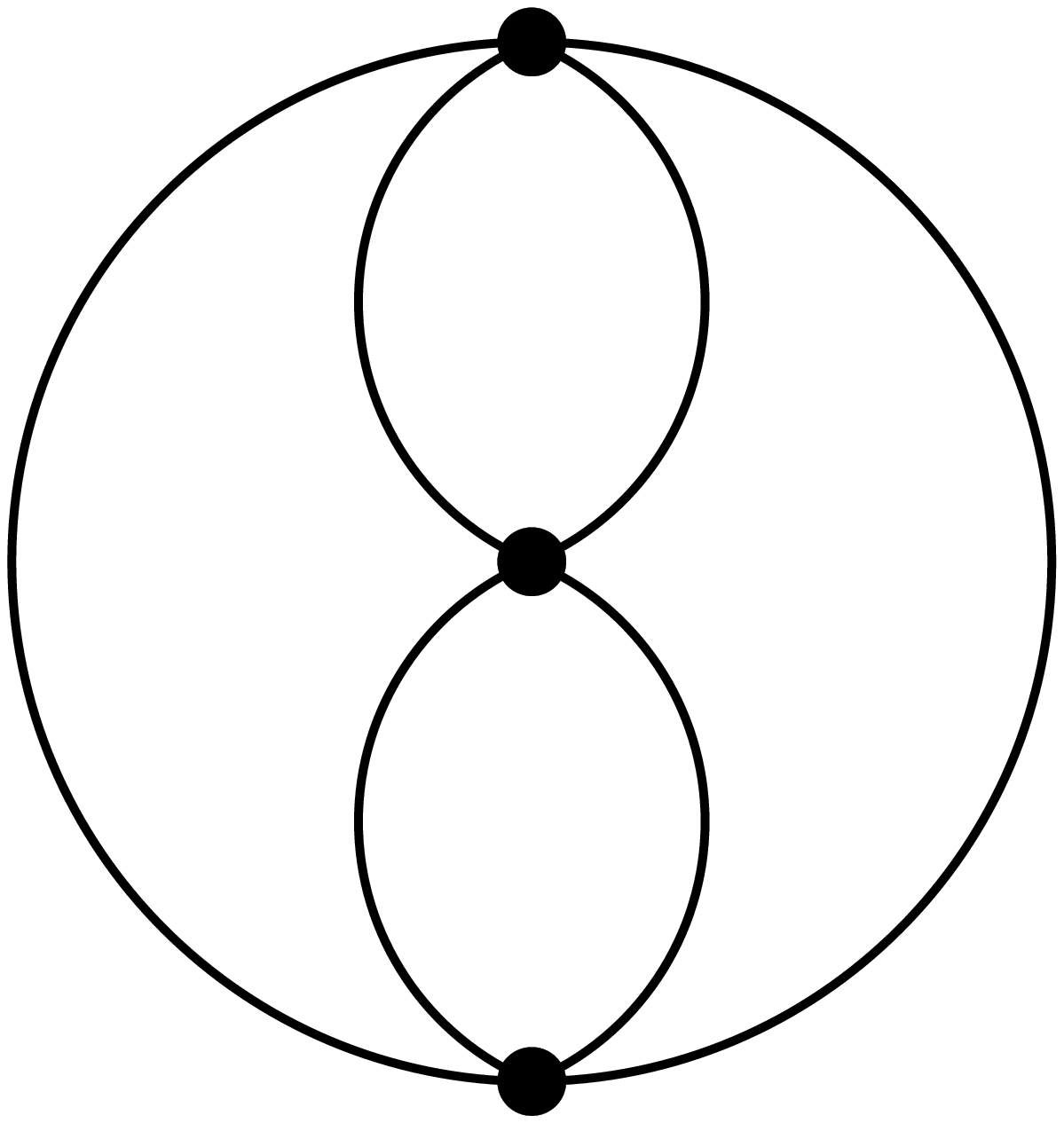}
    $\M{}{6}{1}, (-2,1,1)$
  \end{center}
\end{minipage}
\begin{minipage}[b]{3.0cm}
  \begin{center}
    \includegraphics[height=1.5cm,bb=126 332 460 678]{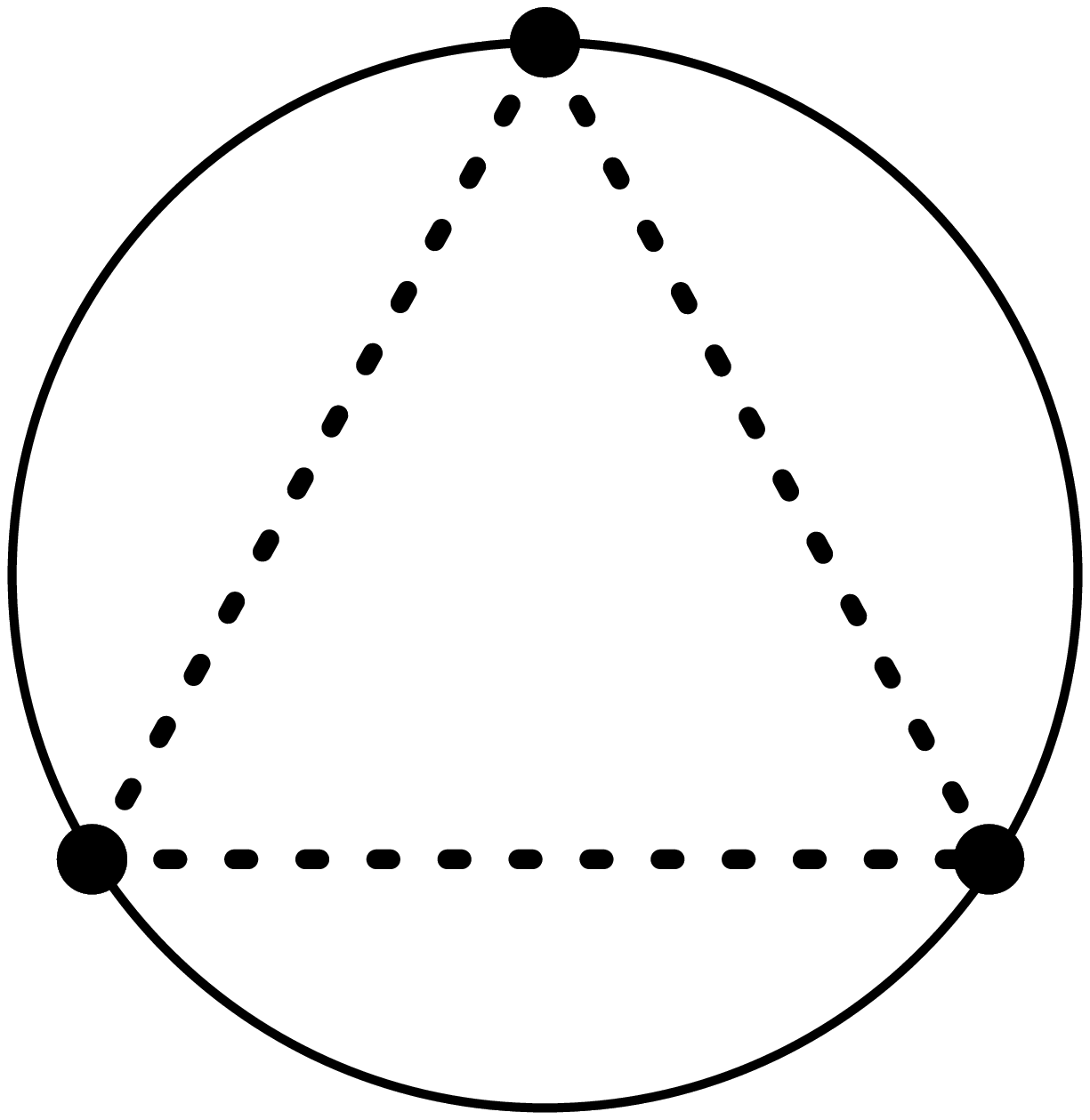}
    $\M{}{6}{4}, (2,1,1)$
  \end{center}
\end{minipage}

\vspace{5mm}

%
\begin{minipage}[b]{3.0cm}
  \begin{center}
    \includegraphics[height=1.5cm,bb=126 332 460 678]{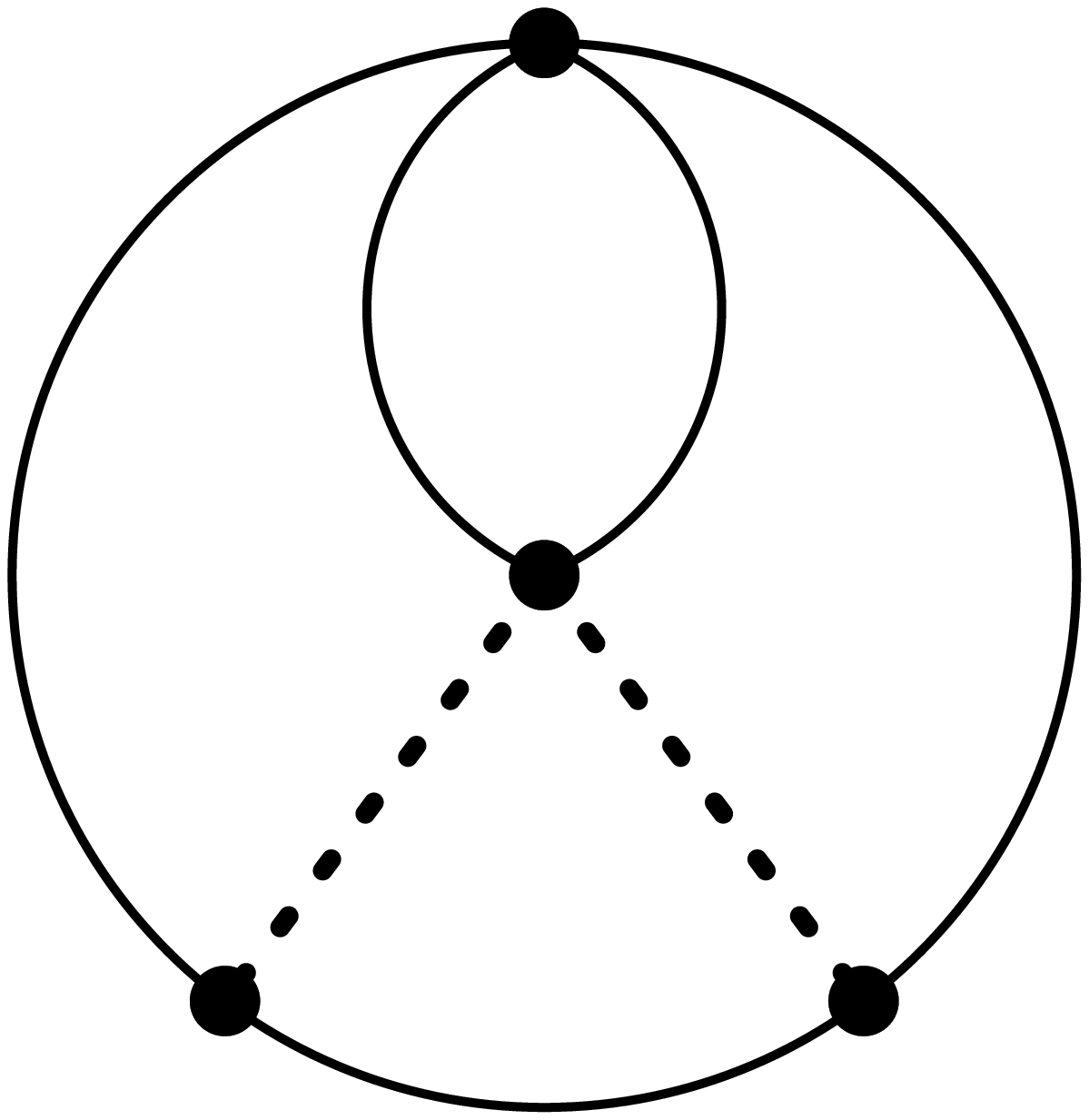}
    $\M{}{7}{1}, (1,0,0)$
  \end{center}
\end{minipage}
%
%
\begin{minipage}[b]{3.0cm}
  \begin{center}
    \includegraphics[height=1.5cm,bb=126 332 460 666]{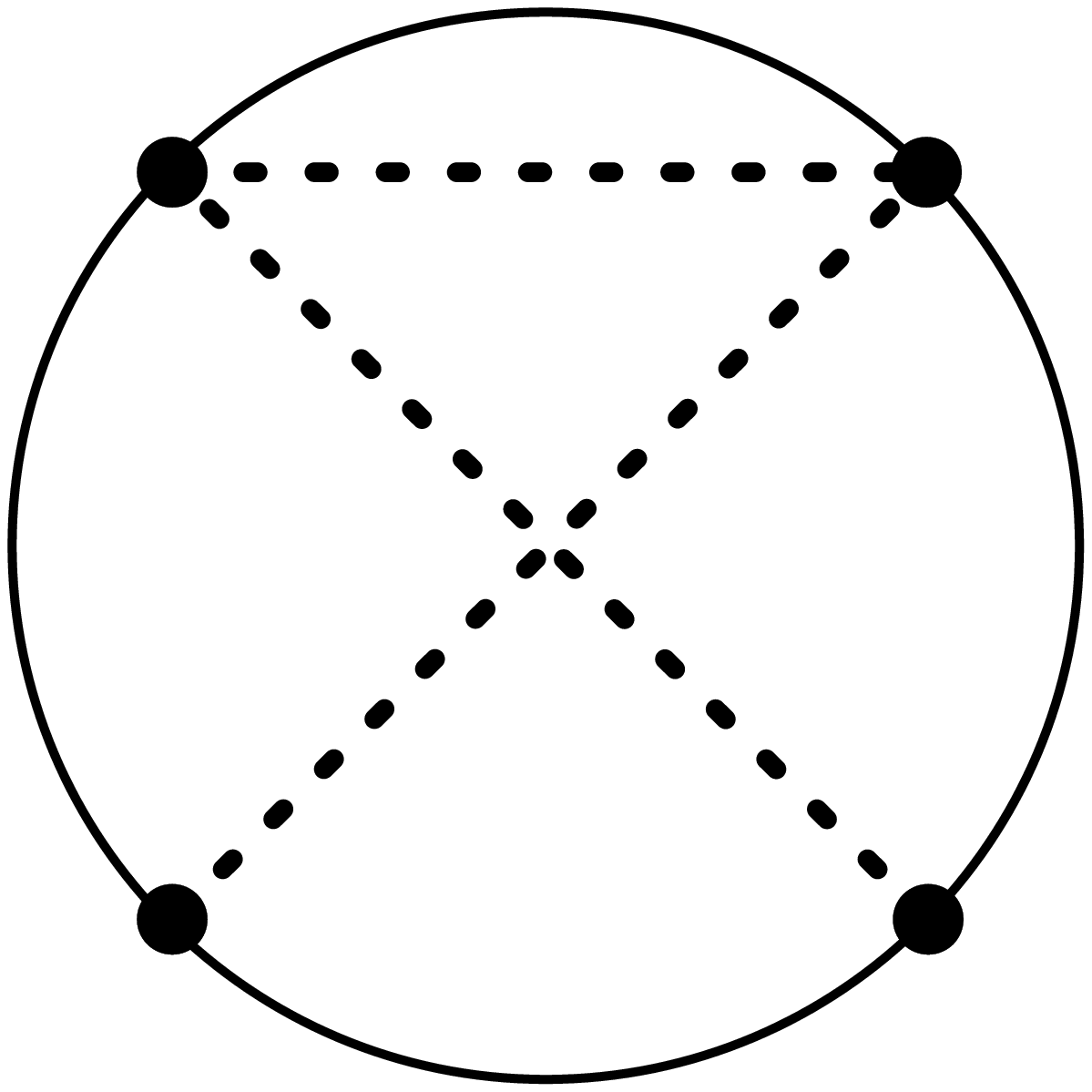}
    $\M{}{7}{2}, (1,0,0)$
  \end{center}
\end{minipage}
%
%
\begin{minipage}[b]{3.0cm}
  \begin{center}
    \includegraphics[height=1.5cm,bb=126 332 460 666]{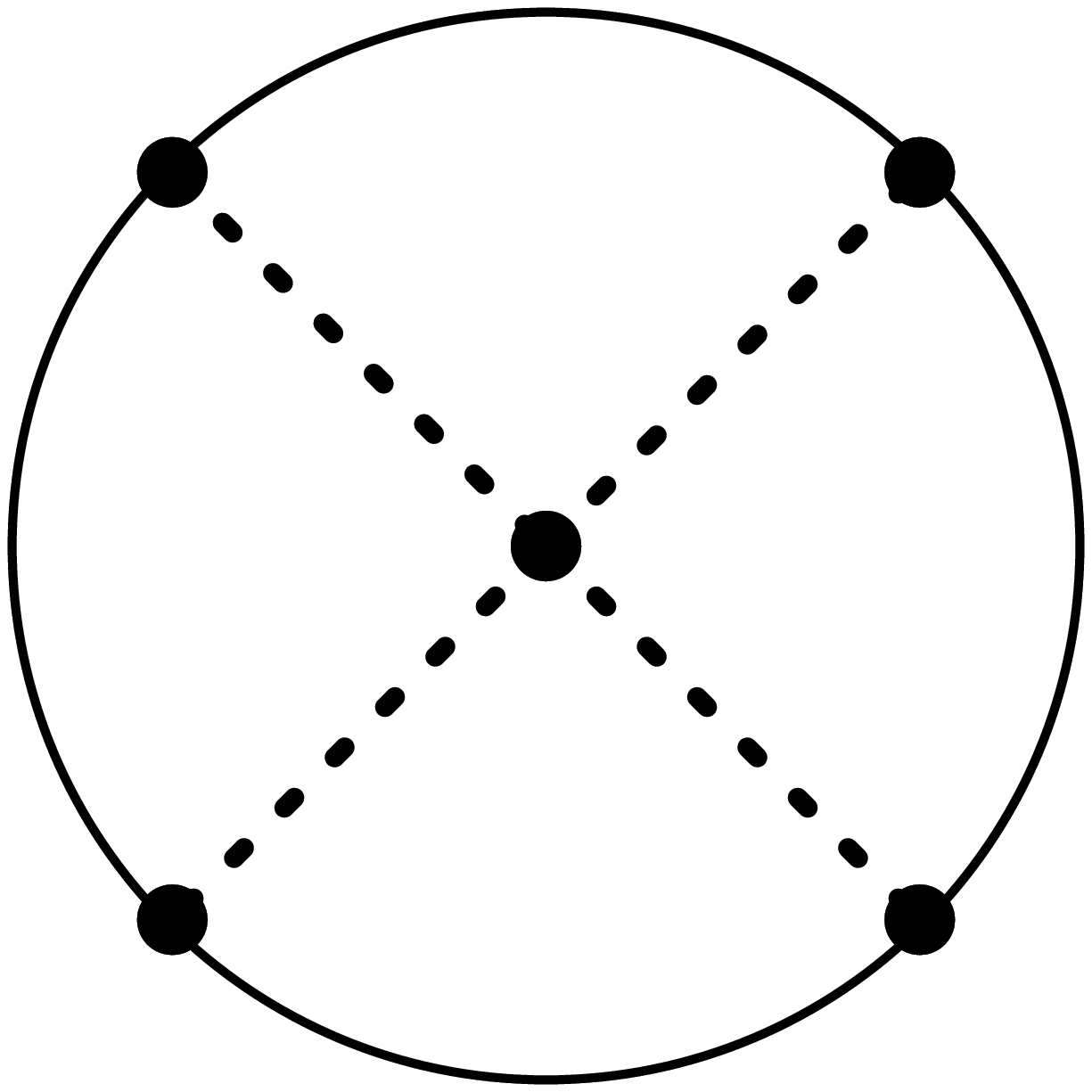}
    $\M{}{8}{1}, (0,-1,-1)$
  \end{center}
\end{minipage}
%
%
\begin{minipage}[b]{3.0cm}
  \begin{center}
    \includegraphics[height=1.5cm,bb=126 320 460 678]{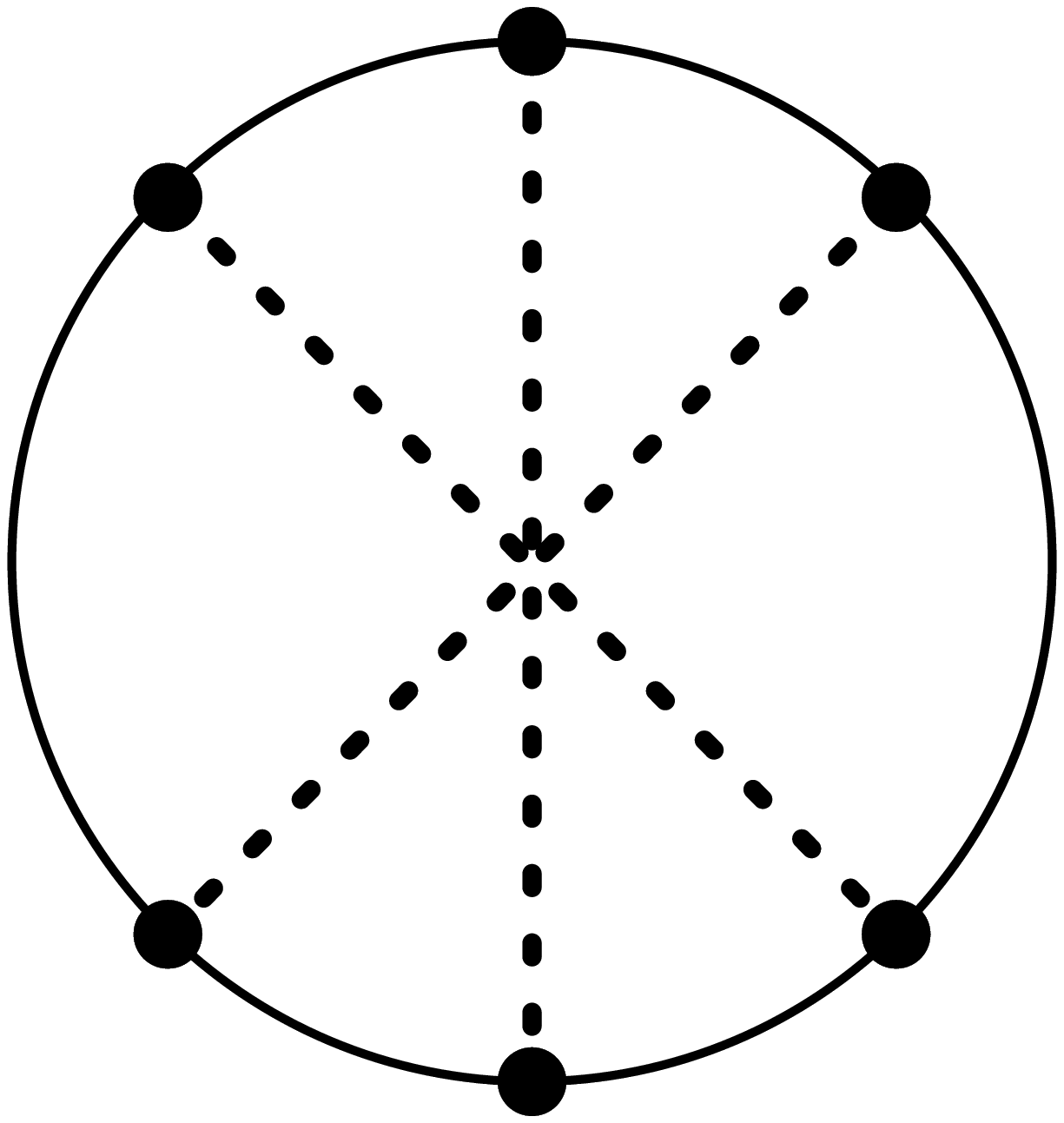}
    $\M{}{9}{1}, (1,0,-1)$
  \end{center}
\end{minipage}
\end{center}

\caption{Master integrals where 
only a few terms  of their $\ep$-expansion are  known
analytically. The solid (dashed) lines denote massive (massless) 
propagators. The three numbers in brackets
$(n_1,n_2,n_3)$ are decoded as follows: $n_1$ is the maximal power of
the spurious pole in $\ep$ which might appear in front of the
integral, $n_2$ is  the maximal power of the spurious pole in $\ep$
which happens to enter  into the decomposition of the bare decoupling constant
$\zeta_g^0$ in terms of the master integrals, ${n_3}$ is the
maximal analytically known power of the $\ep$-expansion of the same
integral as determined in \protect\cite{Schroder:2005va} and
confirmed  in \cite{Chetyrkin:2005}.
\label{MasterTopologies}} 
\end{figure}


\section{Decoupling  at four loops: results \label{sec:res}}

We have generated the relevant diagrams with the help of the program
QGRAF \cite{Nogueira:1991ex}. The total number of  diagrams at four loops
amount to 6070, 765 and 9907 for $\zeta_3^0$, $\tilde\zeta_3^0$ and
$\zeta_1^0$ respectively. All calculations were performed  in a general
covariant gauge with the gluon propagator as given in
eq.~\re{gauge:cov}. However, since extra squared propagators lead to
significant calculational complications,  only linear terms
in $\xi$ were kept.  The bare results are rather lengthy and will be made
available (in  computer-readable form) in
http://www-ttp.physik.uni-karlsruhe.de/Progdata/ttp05/ttp05-27.

After constructing the bare decoupling constant from eq.~\re{zetag0}
followed by the renormalization as described in section~\ref{sec:formalism} we arrive
at the following gauge independent\footnote{In fact, the gauge
dependence on $\xi$ disappears already for the bare decoupling
constant as it should be.} result for the decoupling function $\zeta_g^{\mathrm{MS}}$:
\beq
(\zeta_g^{\mathrm{MS}})^2 = 1 
+ \sum_{i \ge 1} 
\as^i(\mu)
\,
d_{\mathrm{MS,i}}
\label{zetag:ms}
{},
\eeq
where we use the notation
\[
\as(\mu) = \frac{\alpha^{(n_f)}_s(\mu)}{\pi}
\]
and 
the coefficients $d_{\mathrm{MS,i}}$ read

\begin{eqnarray}
d_{\mathrm{MS},1}  =   
-\frac{1}{6} \,\ell_{\mu m}\,
{},
\label{decMS1l}
\end{eqnarray}
\begin{eqnarray}
{d_{\mathrm{MS},2}  =  } 
\frac{11}{72} 
-\frac{11}{24} \,\ell_{\mu m}\,
+\frac{1}{36} \,\ell_{\mu m}^2
{},
\label{decMS2l}
\end{eqnarray}
\begin{eqnarray}
d_{\mathrm{MS},3}  &=  &
\frac{564731}{124416} 
-\frac{82043}{27648}  \,\zeta_{3}
-\frac{955}{576} \,\ell_{\mu m}\,
+\frac{53}{576} \,\ell_{\mu m}^2
-\frac{1}{216} \,\ell_{\mu m}^3
\nonumber\\
&{+}& \, n_l 
\left[
-\frac{2633}{31104} 
+\frac{67}{576} \,\ell_{\mu m}\,
-\frac{1}{36} \,\ell_{\mu m}^2
\right]
{},
\label{decMS3l}
\end{eqnarray}
\begin{eqnarray}
d_{\mathrm{MS},4}  &=&   
 \Delta_{\mathrm{MS},4}\, 
+\frac{7391699}{746496} \,\ell_{\mu m}\,
-\frac{2529743}{165888}  \,\zeta_{3}\,\ell_{\mu m}\,
+\frac{2177}{3456} \,\ell_{\mu m}^2
-\frac{1883}{10368} \,\ell_{\mu m}^3
+\frac{1}{1296} \,\ell_{\mu m}^4
\nonumber\\
&{+}& \, n_l 
\left[
-\frac{110341}{373248} \,\ell_{\mu m}\,
+\frac{110779}{82944}  \,\zeta_{3}\,\ell_{\mu m}\,
-\frac{1483}{10368} \,\ell_{\mu m}^2
-\frac{127}{5184} \,\ell_{\mu m}^3
\right]
\nonumber\\
&{+}& \, n_l^2
\left[
\frac{6865}{186624} \,\ell_{\mu m}\,
-\frac{77}{20736} \,\ell_{\mu m}^2
+\frac{1}{324} \,\ell_{\mu m}^3
\right]
{},
\label{decMS4l}
\end{eqnarray}
\begin{eqnarray}
\lefteqn{\Delta_{\mathrm{MS},4}  =  } 
\nonumber\\
&{}&
\left[
\frac{134805853579559}{43342154956800} 
-\frac{254709337}{783820800} \,\pi^4\,
-\frac{151369}{30481920} \,\pi^6\,
-\frac{18233772727}{783820800}  \,\zeta_{3}
+\frac{151369}{544320}  \,\zeta_3^2
\BreakI
\phantom{+}
+\frac{4330717}{207360}  \,\zeta_{5}
+\frac{9869857}{272160}  \,a_4\,
-\frac{121}{36}  \,a_5\,
-\frac{2057}{51840} \,\pi^4\,\mathrm{ln}\, 2\, 
-\frac{9869857}{6531840} \,\pi^2\,\mathrm{ln}^2\,2\,
\BreakI
\phantom{+}
-\frac{121}{2592} \,\pi^2\,\mathrm{ln}^3\,2\,
+\frac{9869857}{6531840} \mathrm{ln}^4\,2\,
+\frac{121}{4320} \mathrm{ln}^5\,2\,
+\frac{82037}{30965760} \,T_{54,3}\,
-\frac{151369}{11612160} \,T_{62,2}\,
\right]
\nonumber\\
&{+}& \, n_l 
\left[
-\frac{4770941}{2239488} 
-\frac{541549}{14929920} \,\pi^4\,
+\frac{3645913}{995328}  \,\zeta_{3}
+\frac{115}{576}  \,\zeta_{5}
+\frac{685}{5184}  \,a_4\,
\BreakI
\phantom{+ \, n_l }
-\frac{685}{124416} \,\pi^2\,\mathrm{ln}^2\,2\,
+\frac{685}{124416} \mathrm{ln}^4\,2\,
\right]
\nonumber\\
&{+}& \, n_l^2
\left[
-\frac{271883}{4478976} 
+\frac{167}{5184}  \,\zeta_{3}
\right]
{}.
\label{delta4lms}
\end{eqnarray}
In eqs.~(\ref{decMS1l}-\ref{delta4lms}) $\ell_{\mu m} = \ln \frac{\mu^2}{m^2(\mu)}$,
$m(\mu)$ is the (running) heavy quark mass in  the $\msbar$-scheme
and $\mu $ represents the  renormalization  scale. 
Furthermore, $\zeta_n = \zeta(n)$ is Riemann's zeta function and 
$a_n = {\rm Li}_n(1/2) = \sum_{i=1}^{\infty} 1/(2^i i ^n)$.

For another convenient 
definition of the quark mass --- the so-called scale-invariant mass, defined through the
relation  $\mu_h=m(\mu_h)$ ---  eqs.~(\ref{zetag:ms}-\ref{delta4lms}) 
are transformed to 
($\ell_{\mu h} = \ln \frac{\mu^2}{\mu^2_h}$):
\beq
(\zeta_g^{\mathrm{SI}})^2 =
1 
+ \sum_{i \ge 1} 
\as^i(\mu)
\,
d_{\mathrm{SI,i}}
\label{zetag:si}
{},
\eeq
where
\begin{eqnarray}
{d_{\mathrm{SI},1}  =  } 
-\frac{1}{6} \,\ell_{\mu h}\,
{},
\label{decSI1l}
\end{eqnarray}
\begin{eqnarray}
d_{\mathrm{SI},2}  =   
\frac{11}{72} 
-\frac{19}{24} \,\ell_{\mu h}\,
+\frac{1}{36} \,\ell_{\mu h}^2
{},
\label{decSI2l}
\end{eqnarray}
\begin{eqnarray}
d_{\mathrm{SI},3}  &=& 
\frac{564731}{124416} 
-\frac{82043}{27648}  \,\zeta_{3}
-\frac{6793}{1728} \,\ell_{\mu h}\,
-\frac{131}{576} \,\ell_{\mu h}^2
-\frac{1}{216} \,\ell_{\mu h}^3
\nonumber\\
&{+}& \, n_l 
\left[
-\frac{2633}{31104} 
+\frac{281}{1728} \,\ell_{\mu h}\,
\right]
{},
\label{decSI3l}
\end{eqnarray}
\begin{eqnarray}
d_{\mathrm{SI},4}  &=&  
 \Delta_{\mathrm{MS},4}\, 
-\frac{2398621}{746496} \,\ell_{\mu h}\,
-\frac{2483663}{165888}  \,\zeta_{3}\,\ell_{\mu h}\,
-\frac{14023}{3456} \,\ell_{\mu h}^2
-\frac{8371}{10368} \,\ell_{\mu h}^3
+\frac{1}{1296} \,\ell_{\mu h}^4
\nonumber\\
&{+}& \, n_l 
\left[
\frac{190283}{373248} \,\ell_{\mu h}\,
+\frac{133819}{82944}  \,\zeta_{3}\,\ell_{\mu h}\,
+\frac{983}{3456} \,\ell_{\mu h}^2
+\frac{107}{1728} \,\ell_{\mu h}^3
\right]
\nonumber\\
&{+}& \, n_l^2
\left[
\frac{8545}{186624} \,\ell_{\mu h}\,
-\frac{79}{6912} \,\ell_{\mu h}^2
\right]
{}.
\label{decSI4l}
\end{eqnarray}


For practical applications also the inverted  formulae are needed:
\beq
\frac{1}{(\zeta_g^{\mathrm{MS}})^2} =
1
+ \sum_{i \ge 1} 
\left(a^{\prime}_s(\mu)\right)^i 
\,
d^{\prime}_{\mathrm{MS,i}}
\label{unzetag:ms}
{},
\eeq
and
\beq
\frac{1}{(\zeta_g^{\mathrm{SI}})^2} =
1
+ \sum_{i \ge 1} 
\left(a^{\prime}_s(\mu)\right)^i 
\,
d^{\prime}_{\mathrm{SI,i}}
\label{unzetag:si}
{},
\eeq
where we use the notation
\[
a_s^{\prime}(\mu) = \frac{\alpha^{(n_l)}_s(\mu)}{\pi}
\]
and 
the coefficients $d^{\prime}_{\mathrm{MS,i}}$, $d^{\prime}_{\mathrm{SI,i}}$ read
\begin{eqnarray}
d^{\prime}_{\mathrm{MS},1}  =   
\frac{1}{6} \,\ell_{\mu m}\,
{},
\label{undecMS1l}
\end{eqnarray}
\begin{eqnarray}
d^{\prime}_{\mathrm{MS},2}  = 
-\frac{11}{72} 
+\frac{11}{24} \,\ell_{\mu m}\,
+\frac{1}{36} \,\ell_{\mu m}^2
{},
\label{undecMS2l}
\end{eqnarray}
\begin{eqnarray}
d^{\prime}_{\mathrm{MS},3}  = 
&-&\frac{564731}{124416} 
+\frac{82043}{27648}  \,\zeta_{3}
+\frac{2645}{1728} \,\ell_{\mu m}\,
+\frac{167}{576} \,\ell_{\mu m}^2
+\frac{1}{216} \,\ell_{\mu m}^3
\nonumber\\
&{+}& \, n_l 
\left[
\frac{2633}{31104} 
-\frac{67}{576} \,\ell_{\mu m}\,
+\frac{1}{36} \,\ell_{\mu m}^2
\right]
{},
\label{undecMS3l}
\end{eqnarray}
\begin{eqnarray}
\lefteqn{d^{\prime}_{\mathrm{MS},4}  =  } 
\nonumber\\
&{}&
\frac{121}{1728} 
- \Delta_{\mathrm{MS},4}\, 
-\frac{11093717}{746496} \,\ell_{\mu m}\,
+\frac{3022001}{165888}  \,\zeta_{3}\,\ell_{\mu m}\,
+\frac{1837}{1152} \,\ell_{\mu m}^2
+\frac{2909}{10368} \,\ell_{\mu m}^3
+\frac{1}{1296} \,\ell_{\mu m}^4
\nonumber\\
&{+}& \, n_l 
\left[
\frac{141937}{373248} \,\ell_{\mu m}\,
-\frac{110779}{82944}  \,\zeta_{3}\,\ell_{\mu m}\,
+\frac{277}{10368} \,\ell_{\mu m}^2
+\frac{271}{5184} \,\ell_{\mu m}^3
\right]
\nonumber\\
&{+}& \, n_l^2
\left[
-\frac{6865}{186624} \,\ell_{\mu m}\,
+\frac{77}{20736} \,\ell_{\mu m}^2
-\frac{1}{324} \,\ell_{\mu m}^3
\right]
{},
\label{undecMS4l}
\end{eqnarray}

\begin{eqnarray}
d^{\prime}_{\mathrm{SI},1}  = 
\frac{1}{6} \,\ell_{\mu h}\,
{},
\label{undecSI1l}
\end{eqnarray}
\begin{eqnarray}
d^{\prime}_{\mathrm{SI},2}  = 
&{-}&
\frac{11}{72} 
+\frac{19}{24} \,\ell_{\mu h}\,
+\frac{1}{36} \,\ell_{\mu h}^2
{},
\label{undecSI2l}
\end{eqnarray}
\begin{eqnarray}
d^{\prime}_{\mathrm{SI},3}  = 
&-&\frac{564731}{124416} 
+\frac{82043}{27648}  \,\zeta_{3}
+\frac{2191}{576} \,\ell_{\mu h}\,
+\frac{511}{576} \,\ell_{\mu h}^2
+\frac{1}{216} \,\ell_{\mu h}^3
\nonumber\\
&{+}& \, n_l 
\left[
\frac{2633}{31104} 
-\frac{281}{1728} \,\ell_{\mu h}\,
\right]
{},
\label{undecSI3l}
\end{eqnarray}
\begin{eqnarray}
\lefteqn{d^{\prime}_{\mathrm{SI},4}  =  } 
\nonumber\\
&{}&
\frac{121}{1728} 
- \Delta_{\mathrm{MS},4}\, 
-\frac{1531493}{746496} \,\ell_{\mu h}\,
+\frac{2975921}{165888}  \,\zeta_{3}\,\ell_{\mu h}\,
+\frac{33887}{3456} \,\ell_{\mu h}^2
+\frac{14149}{10368} \,\ell_{\mu h}^3
+\frac{1}{1296} \,\ell_{\mu h}^4
\nonumber\\
&{+}& \, n_l 
\left[
-\frac{158687}{373248} \,\ell_{\mu h}\,
-\frac{133819}{82944}  \,\zeta_{3}\,\ell_{\mu h}\,
-\frac{515}{1152} \,\ell_{\mu h}^2
-\frac{107}{1728} \,\ell_{\mu h}^3
\right]
\nonumber\\
&{+}& \, n_l^2
\left[
-\frac{8545}{186624} \,\ell_{\mu h}\,
+\frac{79}{6912} \,\ell_{\mu h}^2
\right]
{}.
\label{undecSI4l}
\end{eqnarray}
Numerically, eqs.~\re{zetag:ms} and \re{zetag:si} read
\bea
(\zeta_g^{\mathrm{MS}})^2 
\,\,\bbuildrel{=\!=\!=}_{\mu=\mu_h}^{} \,\, 
1
&\n+\n&  0.152778\,{\as^2(\mu_h)} + 
 {\as^3(\mu_h)}\,\left( 0.972057 - 0.0846515\,\,n_l \right)  
\nonumber
\\
&{+}&
 {\as^4(\mu_h)}\,\left( 5.17035 - 1.00993\,\,n_l - 
    0.0219784\,{\,n_l}^2 \right) 
{},
\label{decMSnum}
\eea
\bea
\frac{1}{(\zeta_g^{\mathrm{MS}})^2} 
\,\,\bbuildrel{=\!=\!=}_{\mu=\mu_h}^{} \,\,
1
&\n-\n&
 0.152778\,{\as^{\prime \, 2}(\mu_h)} + 
 {\as^{\prime \, 3}(\mu_h)}\,\left( -0.972057 + 0.0846515\,\,n_l \right)  
\nonumber
\\
&{+}&
 {\as^{\prime \, 4}(\mu_h)}\,\left( -5.10032 + 1.00993\,\,n_l + 
    0.0219784\,{\,n_l}^2 \right) 
\label{unddecMSnum}
{}.
\eea

Using the three loop relation between the pole and $\msbar$ masses 
\cite{Chetyrkin:1999ys,Chetyrkin:1999qi,Melnikov:2000qh} one could 
easily express the decoupling relations in terms of the
pole mass of the heavy quark. As the resulting expressions are rather lengthy
they   are  relegated to the Appendix.

\section{\ice{Effective} Coupling of the Higgs boson to gluons \label{sec:C1}}

Within the Standard Model (SM) the scalar Higgs boson is responsible
for the mechanism of the mass generation.  
Its future \mbox{(non-)discovery} will be of primary importance for all
the particle physics.  The SM Higgs boson mass is constrained from
below, $M_h > 114 \GeV$, by experiments at LEP and SLC
\cite{Barate:2003sz,:2004qh}.  Indirect
constraints coming from precision electroweak measurements
\cite{Carena:2002es} lead to an upper limit of about  200 $\GeV$.

With the SM Higgs boson mass within this range, its coupling to a pair
of gluons is mediated by virtual quarks \cite{Wilczek:1977zn} and
plays a crucial r\^ole in Higgs phenomenology.  Indeed, with Yukawa
couplings of the Higgs boson to  quarks  being proportional to
the respective quark masses, the $ggH$ coupling of the SM is
essentially generated by the top quark only.  The $ggH$ coupling
strength becomes independent of the top quark mass $M_t$ in the limit
$M_H\ll 2M_t$.

In general, the theoretical description of such interactions is very
complicated because there are two different mass scales involved, $M_H$ and
$M_t$. However, in the limit $M_H\ll2M_t$, the situation may be greatly simplified by
integrating out the top quark, i.e. by constructing a heavy-top-quark
effective Lagrangian ~\cite{Inami:1982xt,Chetyrkin:1997iv}.

This Lagrangian is a linear combination of certain dimension-four
operators acting in QCD with five quark flavors, while all $M_t$
dependence is contained in the coefficient functions.  
The final renormalized version of ${\cal L}_{\rm eff}$ is
($G_F$ is Fermi's constant)
\begin{equation}
\label{eff}
{\cal L}_{\rm eff}=-2^{1/4}G_F^{1/2}HC_1\left[O_1^\prime\right]
{}.
\end{equation}
Here, $\left[O_1^\prime\right]$ is the renormalized counterpart of the bare
operator $O_1^\prime=G_{a\mu\nu}^{0\prime}G_a^{0\prime\mu\nu}$, where
$G_{a\mu\nu}$ is the color field strength, the superscript 0 denotes bare
fields, and primed objects refer to the five-flavor effective theory.
$C_1$ is the corresponding renormalized coefficient function, which carries
all $M_t$ dependence.

In  ref.~\cite{Chetyrkin:1998un}
a low-energy theorem was established, valid to all orders, which relates the
effective coupling of the Higgs boson to gluons, induced by a 
presence of  heavy quark $h$, to the logarithmic derivative of $\zeta_g$  w.r.t.\
$m$. The theorem states that
\beq
C_1 \,=\, -\frac{1}{2}\,  m^2\, \frac{\prt}{\prt \,m^2}\ln\zeta_g^2
\label{let1}
{}.
\eeq
Another, equivalent, but more convenient form of  \re{let1} was also derived 
in \cite{Chetyrkin:1998un}  by exploiting  evolution equations in  full and effective
theories. It reads
\begin{equation}
C_1=\frac{\pi}{2\left[1-2\gamma_m(\as)\right]}
\left[\hat{\beta}^\prime(\as^\prime)
-\hat{\beta}(\as)
-\hat{\beta} (\as)\, \as \frac{\partial}{\partial\as}
 \ln\zeta_g^2
\right]
\label{let1:rg}
{},
\end{equation}
where $\g_m$ is  the quark mass anomalous dimension, 
$\zeta_g^2=\zeta_g^2(\mu,\as,m)$ and the
$\beta$-function is defined as follows:
\beq
\mu^2 \frac{d}{d\, \mu^2}\,a_s^{(n_f)}\,=\,
\beta^{(n_f)}\left(a_s^{(n_f)}\right)
= 
-\sum_{N=1}^\infty\beta_{N-1}^{(n_f)}
a_s^{N+1}
\label{eqrgea}
{},
\eeq
\[
\as=\as^{(n_f)},
\ \ 
 \as^{\prime} =  \as^{(n_l)},
 \ \ 
\as\,\hat{\beta}(\as) = \beta^{(n_f)}(\as),
\ \ 
a_s^{\prime}\, \hat{\beta}^{\prime}(a_s^{\prime}) = \beta^{(n_l)}(a_s^{\prime})
{}.
\]

An important feature of the low energy theorem \re{let1:rg} is the
fact that the decoupling function $\zeta_g$ appears there multiplied
by at least one power of $\as$.  It means that in order to compute
$C_1$ , say, at four loops one should know the QCD $\beta$-function to
{\em four loops} ( only $n_f$ dependent part) but the quark anomalous
dimension $\g_m$ and the decoupling function $\zeta_g$ only to {\em
three loops}. It is this fact which  allowed to find $C_1$ at four loops
in \cite{Chetyrkin:1998un} long before   
 the {\em  four-loop} result for the decoupling
function would be available. 

Now, armed with the newly computed four-loop term in $\zeta_g$, we
could easily check that the old result for $C_1$ of
\cite{Chetyrkin:1998un} by a direct use of eq.~\re{let1}. We have done
this simple exercise and found full agreement. In fact, one could even
use eq.~\re{let1:rg} in order to construct $C_1$ at five loops in
terms of the known  four-loop QCD decoupling function, the quark anomalous
dimension $\g_m$ and the $\beta$-function and the only yet unknown
parameter: the $n_f$-dependent piece of the five-loop contribution to
the  $\beta$-function. The result reads
\begin{eqnarray}
&\n{}\n& C_1=
-\frac{1}{12}\,\as(\mu_h)
\Bigg\{
  1 
+ \frac{11}{4}\, \as(\mu_h) 
{}+ \as^2(\mu_h)
\Bigg[
\frac{2821}{288} 
+ n_l\left(
-\frac{67}{96} 
\right)
\Bigg]
\nonumber\\
&\n+\n&{} \as^3(\mu_h)
\Bigg[
-\frac{4004351}{62208} 
+ \frac{1305893}{13824}\zeta(3)
+  n_l \left(
  \frac{115607}{62208} 
- \frac{110779}{13824}\zeta(3)
\right) 
{}+ n_l^2 \left(
- \frac{6865}{31104} 
\right)
\Bigg]
\nonumber
\\
&\n+\n& \as^4(\mu_h)
\Bigg[
-\frac{13546105}{41472}
-\frac{91}{2} \Delta_{\mathrm{MS},4}\,
-\frac{31}{432} \,\pi^4\,
-\frac{853091}{6912}  \,\zeta_{3}
+\frac{475}{9}  \,\zeta_{5}
\nonumber\\
&{+}& \, n_l
\Bigg(
\frac{28598581}{497664}
 + 3  \Delta_{\mathrm{MS},4}\,
-\frac{29}{432} \,\pi^4\,
+\frac{3843215}{110592}  \,\zeta_{3}
-\frac{575}{36}  \,\zeta_{5}
\Bigg)
\nonumber\\
&{+}& \, n_l^2
\bigg(
-\frac{3503}{62208}
+\frac{1}{216} \,\pi^4\,
-\frac{3}{4}  \,\zeta_{3}
\Bigg)
{+} \, n_l^3
\Bigg(
\frac{83}{7776}
-\frac{1}{54}  \,\zeta_{3}
\Bigg)
+ 6\, \Delta \beta_4 
\Bigg]
\Bigg\}
\nonumber
\\
&\n\approx\n&
-\frac{1}{12}\,\as(\mu_h)
\Bigg[1
+ 2.7500
\, \as(\mu_h)
+ \left(9.7951 - 0.6979\,n_l\right) 
\as^2(\mu_h)
\nonumber\\
&\n\n&
+ \left(49.1827 - 7.7743\,n_l - 0.2207\,n_l^2 \right)
\as^3(\mu_h)
\nonumber\\
&\n\n&
+
\left(
-662.507 + 137.601\, n_l - 2.53666 \,n^2_l 
 - 0.077522\,  n^3_l 
+ 6\, \Delta \beta_4 
\right)
\as^4(\mu_h)
\Bigg],
\label{C1:5loop}
\eea
 \ice{
light-beta approx. for MSBAR mass max loop = 5
                 2
1. + 2.75 as + as  (9.79514 - 0.697917 nl) + 
   3                                    2
 as  (49.1827 - 7.77432 nl - 0.220711 nl ) + 
   4                                    2               3
 as  (-662.507 + 137.601 nl - 2.53666 nl  - 0.0775216 nl )
 }
where 
\[
\Delta \beta_4 =   \beta^{(n_f-1)}_4 - \beta^{(n_f)}_4
{}.
\]

To save space we have written eq.~\re{C1:5loop} for the value of $\mu =\mu_h$.
Unfortunately,   simple estimations show 
that the $\beta$-function dependent contribution 
to eq.~\re{C1:5loop}  could be numerically important.

\section{Application: $\alpha_s(M_Z)$ from  $\alpha_s(M_{\tau})$ \label{sec:tau}}

A central feature of QCD is the possibility to describe measurements
performed at  very different energy scales with the help of only one
coupling constant.  The customary choice is the ($\msbar$) coupling
constant $\alpha_s(M_Z)$ which can be measured precisely in Z-boson
decays. (For a review see, e.g.  \cite{ChKK:Report:1996}.)

On the other hand the dependence of the $\tau$-decay rate on the
strong coupling $\alpha_s$ has been used for a determination of $\als$
at lower energies, with the most recent results 
of $0.340 \pm 0.005_\mathrm{exp} \pm 0.014_\mathrm{th}$ 
and $0.348\pm
0.009_\mathrm{exp} \pm 0.019_\mathrm{th }$ by the ALEPH \cite{Schael:2005am}
and OPAL \cite{Ackerstaff:1998yj} collaborations.

After evolution up to higher energies (taking into account the
threshold effects due to the $c$- and $b$-quarks) these results agree
remarkably well with determinations based on the hadronic $Z$ decay
rate, which provides a genuine test of asymptotic freedom of QCD.

Very recently an  analysis of the evolution has been performed in \cite{Davier:2005xq}.
They start from an updated determination of $\alpha_s(m_{\tau})$ 
\beq
\alpha_s(m_{\tau}) = 0.345 \pm 0.004_{\rm exp} \pm 0.008_{\rm th}
\label{astau:final}
{}
\eeq
based on
most recent experimental results of the ALEPH collaboration \cite{Schael:2005am}.
Their result for the  evolution of  $\asm$ given in eq.~(\ref{astau:final}), 
based on 
the use of the four-loop running and  and three-loop quark-flavor
matching reads
\bea
\label{eq:asres_mz}
   \as(M_Z) &=& 0.1215 \pm 0.0004_{\rm exp}
                                       \pm 0.0010_{\rm th}
                                       \pm 0.0005_{\rm evol}~, \nonumber \\
                         &=& 0.1215 \pm 0.0012~
{},
\eea
where the first two errors originate from the $\asm$ determination
given in Eq.~(\ref{astau:final}), and the last error stands for the
ambiguities in the evolution due to uncertainties in the matching
scales of the quark thresholds. That evolution error received
contributions from the uncertainties in the $c$-quark mass (0.00020,
$\mu_c=1.31 \pm 0.1$ GeV) and the $b$-quark mass (0.00005, $\mu_b=
4.13\pm 0.1$ GeV), the matching scale (0.00023, $\mu$ varied between
$0.7\,\mu_q$ and $3.0\,\mu_q$), the three-loop truncation in the
matching expansion (0.00026) and the four-loop truncation in the RGE
equation (0.00031). (For the last two errors the size of the shift due
the highest known perturbative term was treated as systematic
uncertainty.)  The errors had been added in quadrature.

We have repeated this analysis including the newly computed 
four-loop approximation for the matching.\footnote{We have used the 
package RunDec \cite{Chetyrkin:2000yt}.}
As a result the value of 
$\as(M_Z)$ has been marginally increased (by 0.0001) and both errors 
from the choice of the matching scale and from  
the four-loop truncation in the matching equation have been halved.
The updated  version of \re{eq:asres_mz} now reads
\bea
\label{eq:asres_mz:new}
   \as(M_Z) &=& 0.1216 \pm 0.0004_{\rm exp}
                                       \pm 0.0010_{\rm th}
                                       \pm 0.0004_{\rm evol}~, \nonumber \\
                         &=& 0.1216 \pm 0.0012~
{}.
\eea

\section{Summary and Conclusions \label{sec:conclusions} }

We have computed the decoupling relation for the QCD quark gluon
coupling constant in four-loop approximation. The new contribution
leads to a decrease of the matching related
uncertainties in the process of the evolution of the
$\alpha_s(m_{\tau})$ to $\alpha_s(M_{Z})$ by a factor of two.

As a by-product we have directly confirmed the long available result
for $C_1$, the effective coupling of the Higgs boson to gluons at four loops, 
and (partially) extended it to five loops. It remains only to find the QCD 
$\beta$-function at {\em five} loops to  get the full result for $C_1$.
In the light of recent significant  progress in the multiloop technology
\cite{Baikov:2004tk,Baikov:2005rw} the completely analytical evaluation 
of the  former seems to be  possible in not too far distant future.

We would like to mention that the result for the decoupling function
at four loops  as well as for the   Higgs boson to gluons effective coupling at 
five loops have been obtained by a completely independent calculation in 
\cite{YS_MS}. The results are in full  mutual agreement. 
\vspace{1cm}

\noindent
{\bf Acknowledgments}
\vspace{1cm}

The authors are grateful to York Schr\"oder and Matthias  Steinhauser 
for useful discussions and information about \cite{YS_MS}.
K. Ch. thanks Michail Kalmykov for some  useful  advice about 
integral $T_{52}$.

The work was supported by the Deutsche Forschungsgemeinschaft in the
Sonderforschungsbereich/Transregio SFB/TR-9 ``Computational Particle
Physics''.  The work of C.S. was also partially supported by MIUR
under contract 2001023713$\_$006.

\appendix
\section{
\label{app:os} $\zeta_g$ for the heavy quark mass  in the on-shell scheme}


The relevant formulas for the case of the heavy quark mass 
renormalized in the on-shell scheme  (denoted as $M$) are:
\beq
(\zeta_g^{\mathrm{OS}})^2 = 1 
+ \sum_{i \ge 1} 
\as^i(\mu)
\,
d_{\mathrm{OS,i}}
\label{zetag:os}
{}
\eeq
and 
\beq
\frac{1}{(\zeta_g^{\mathrm{OS}})^2} = 1 
+ \sum_{i \ge 1} 
\asp{i}(\mu)
\,
d^{\prime}_{\mathrm{OS,i}}
\label{unzetag:os}
{},
\eeq
where
$\ell_{\mu M} = \ln \frac{\mu^2}{M^2}$ and
\begin{eqnarray}
{d_{\mathrm{OS},1}  =  }
-\frac{1}{6} \,\ell_{\mu M}\,
{},
\label{decOS1l}
\end{eqnarray}
\begin{eqnarray}
d_{\mathrm{OS},2}  =   
-\frac{7}{24} 
-\frac{19}{24} \,\ell_{\mu M}\,
+\frac{1}{36} \,\ell_{\mu M}^2
{},
\label{decOS2l}
\end{eqnarray}
\begin{eqnarray}
d_{\mathrm{OS},3}  &=& 
-\frac{58933}{124416} 
-\frac{1}{9} \,\pi^2\,
-\frac{80507}{27648}  \,\zeta_{3}
-\frac{1}{27} \,\pi^2\,\mathrm{ln}\, 2\, 
-\frac{8521}{1728} \,\ell_{\mu M}\,
\nonumber
\\
&-&\frac{131}{576} \,\ell_{\mu M}^2
-\frac{1}{216} \,\ell_{\mu M}^3
%
+ \, n_l 
\left[
\frac{2479}{31104} 
+\frac{1}{54} \,\pi^2\,
+\frac{409}{1728} \,\ell_{\mu M}\,
\right]
{},
\label{decOS3l}
\end{eqnarray}
\begin{eqnarray}
d_{\mathrm{OS},4}  &=&  
 \Delta_{\mathrm{OS},4}\, 
-\frac{19696909}{746496} \,\ell_{\mu M}\,
-\frac{29}{54} \,\pi^2\,\,\ell_{\mu M}\,
-\frac{2439119}{165888}  \,\zeta_{3}\,\ell_{\mu M}\,
-\frac{29}{162} \,\pi^2\,\mathrm{ln}\, 2\, \,\ell_{\mu M}\,
\nonumber
\\
&-&\frac{7693}{1152} \,\ell_{\mu M}^2
-\frac{8371}{10368} \,\ell_{\mu M}^3
+\frac{1}{1296} \,\ell_{\mu M}^4
\nonumber\\
&{+}& \, n_l 
\left[
\frac{1110443}{373248} \,\ell_{\mu M}\,
+\frac{41}{324} \,\pi^2\,\,\ell_{\mu M}\,
+\frac{132283}{82944}  \,\zeta_{3}\,\ell_{\mu M}\,
+\frac{1}{81} \,\pi^2\,\mathrm{ln}\, 2\, \,\ell_{\mu M}\,
+\frac{6661}{10368} \,\ell_{\mu M}^2
\BreakI
\phantom{+ \, n_l }
+\frac{107}{1728} \,\ell_{\mu M}^3
\right]
\nonumber\\
&{+}& \, n_l^2
\left[
-\frac{1679}{186624} \,\ell_{\mu M}\,
-\frac{1}{162} \,\pi^2\,\,\ell_{\mu M}\,
-\frac{493}{20736} \,\ell_{\mu M}^2
\right]
{},
\label{decOS4l}
\end{eqnarray}
\begin{eqnarray}
\lefteqn{\Delta_{\mathrm{OS},4}  =  } 
\nonumber\\
&-&
\left.
\frac{2180918653146841}{43342154956800} 
-\frac{697121}{116640} \,\pi^2\,
-\frac{231357337}{783820800} \,\pi^4\,
-\frac{151369}{30481920} \,\pi^6\,
-\frac{18646246327}{783820800}  \,\zeta_{3}
\BreakI
\phantom{+}
+\frac{1439}{1296} \,\pi^2\, \,\zeta_{3}
+\frac{151369}{544320}  \,\zeta_3^2
+\frac{3698717}{207360}  \,\zeta_{5}
+\frac{10609057}{272160}  \,a_4\,
-\frac{121}{36}  \,a_5\,
\BreakI
\phantom{+}
+\frac{1027}{972} \,\pi^2\,\mathrm{ln}\, 2\, 
-\frac{2057}{51840} \,\pi^4\,\mathrm{ln}\, 2\, 
-\frac{9278497}{6531840} \,\pi^2\,\mathrm{ln}^2\,2\,
-\frac{121}{2592} \,\pi^2\,\mathrm{ln}^3\,2\,
+\frac{10609057}{6531840} \mathrm{ln}^4\,2\,
\BreakI
\phantom{+}
+\frac{121}{4320} \mathrm{ln}^5\,2\,
+\frac{82037}{30965760} \,T_{54,3}\,
-\frac{151369}{11612160} \,T_{62,2}\,
\right.
\nonumber\\
&{+}& \, n_l 
\left[
\frac{1773073}{746496} 
+\frac{557}{972} \,\pi^2\,
-\frac{697709}{14929920} \,\pi^4\,
+\frac{4756441}{995328}  \,\zeta_{3}
+\frac{115}{576}  \,\zeta_{5}
\BreakI
\phantom{+ \, n_l }
+\frac{173}{5184}  \,a_4\,
+\frac{11}{243} \,\pi^2\,\mathrm{ln}\, 2\, 
-\frac{1709}{124416} \,\pi^2\,\mathrm{ln}^2\,2\,
+\frac{173}{124416} \mathrm{ln}^4\,2\,
\right]
\nonumber\\
&{+}& \, n_l^2
\left[
-\frac{140825}{1492992} 
-\frac{13}{972} \,\pi^2\,
-\frac{19}{1728}  \,\zeta_{3}
\right]
{},
\label{delta4los}
\end{eqnarray}
\begin{eqnarray}
{d^{\prime}_{\mathrm{OS},1}  =  }
\frac{1}{6} \,\ell_{\mu M}\,
{},
\label{undecOS1l}
\end{eqnarray}
\begin{eqnarray}
  d^{\prime}_{\mathrm{OS},2}  =  
 \frac{7}{24} 
+\frac{19}{24} \,\ell_{\mu M}\,
+\frac{1}{36} \,\ell_{\mu M}^2
{},
\label{undecOS2l}
\end{eqnarray}
\begin{eqnarray}
d^{\prime}_{\mathrm{OS},3}  = &{}& 
\frac{58933}{124416} 
+\frac{1}{9} \,\pi^2\,
+\frac{80507}{27648}  \,\zeta_{3}
+\frac{1}{27} \,\pi^2\,\mathrm{ln}\, 2\, 
+\frac{8941}{1728} \,\ell_{\mu M}\,
\nonumber
\\
&+&\frac{511}{576} \,\ell_{\mu M}^2
+\frac{1}{216} \,\ell_{\mu M}^3
\nonumber\\
&{+}& \, n_l 
\left[
-\frac{2479}{31104} 
-\frac{1}{54} \,\pi^2\,
-\frac{409}{1728} \,\ell_{\mu M}\,
\right]
{},
\label{undecOS3l}
\end{eqnarray}
\begin{eqnarray}
d^{\prime}_{\mathrm{OS},4}  =  
&{}&
\left.
\frac{49}{192} 
- \Delta_{\mathrm{OS},4}\, 
+\frac{21084715}{746496} \,\ell_{\mu M}\,
+\frac{35}{54} \,\pi^2\,\,\ell_{\mu M}\,
+\frac{2922161}{165888}  \,\zeta_{3}\,\ell_{\mu M}\,
\BreakI
\phantom{+}
+\frac{35}{162} \,\pi^2\,\mathrm{ln}\, 2\, \,\ell_{\mu M}\,
+\frac{47039}{3456} \,\ell_{\mu M}^2
+\frac{14149}{10368} \,\ell_{\mu M}^3
+\frac{1}{1296} \,\ell_{\mu M}^4
\right.
\nonumber\\
&{+}& \, n_l 
\left[
-\frac{1140191}{373248} \,\ell_{\mu M}\,
-\frac{47}{324} \,\pi^2\,\,\ell_{\mu M}\,
-\frac{132283}{82944}  \,\zeta_{3}\,\ell_{\mu M}\,
-\frac{1}{81} \,\pi^2\,\mathrm{ln}\, 2\, \,\ell_{\mu M}\,
-\frac{9115}{10368} \,\ell_{\mu M}^2
\BreakI
\phantom{+ \, n_l }
-\frac{107}{1728} \,\ell_{\mu M}^3
\right]
\nonumber\\
&{+}& \, n_l^2
\left[
\frac{1679}{186624} \,\ell_{\mu M}\,
+\frac{1}{162} \,\pi^2\,\,\ell_{\mu M}\,
+\frac{493}{20736} \,\ell_{\mu M}^2
\right]
{},
\label{undecOS4l}
\end{eqnarray}
\bea
(\zeta_g^{\mathrm{OS}})^2 
\,\,\bbuildrel{=\!=\!=}_{\mu=M}^{} \,\,  
1
&\n-\n& 
 0.291667\,{\as^2(M)} + 
 {\as^3(M)}\,\left( -5.32389 + 0.262471\,\,n_l \right)  
\nonumber
\\
&{+}&
 {\as^4(M)}\,\left( -85.875 + 9.69229\,\,n_l - 
    0.239542\,{\,n_l}^2 \right) 
{},
\label{decOSnum}
\eea
\bea
\frac{1}{(\zeta_g^{\mathrm{OS}})^2}
\,\,\bbuildrel{=\!=\!=}_{\mu=M}^{} \,\,  
1
&\n+\n&
0.291667\,{\as^{\prime \, 2 }(M)} + 
 {\as^{\prime \, 3}(M)}\,\left( 5.32389 - 0.262471\,\,n_l \right)  
\nonumber
\\
&{+}&
 {\as^{\prime \, 4}(M) }\,\left( 86.1302 - 9.69229\,\,n_l + 
    0.239542\,{\,n_l}^2 \right) 
{}.
\eea

\end{document}